\documentclass[10pt,sigconf]{acmart}
\renewcommand\footnotetextcopyrightpermission[1]{} 
\usepackage{adjustbox}

\usepackage{tablefootnote}

\usepackage{tikzsymbols}
\usepackage{pifont}
\newcommand{\cmark}{\ding{51}}%
\newcommand{\xmark}{\ding{55}}%

\usepackage{xcolor}
\usepackage{colortbl}%
  \newcommand{\grayrow}{\rowcolor[gray]{0.925}}
\usepackage{booktabs}
\usepackage{wrapfig}

\newcommand\ml{ML\xspace}
\newcommand\name{IIsy\xspace}

\newcommand\dc{data center\xspace}

\newcommand\dcs{data centers\xspace}

\newcommand\inc{in-network computing\xspace}
\newcommand\inl{in-network classification\xspace}
\newcommand\Inl{In-network classification\xspace}
\newcommand\INL{In-Network Classification\xspace}
\newcommand\lut{lookup tables\xspace}

\usepackage{hyperref}
\usepackage{comment}
\usepackage{enumitem}
\setlist[itemize]{noitemsep, nolistsep}
\hypersetup{pdfstartview=FitH,pdfpagelayout=SinglePage}

\usepackage{xspace}
\usepackage{ifthen}
\usepackage{flushend}
\usepackage{color}
\usepackage{booktabs}
\newboolean{draft}
\setboolean{draft}{false}
\newboolean{comments}
\ifdraft
\setboolean{comments}{true}
\else
\setboolean{comments}{false}
\fi

\renewcommand\footnotetextcopyrightpermission[1]{} 
\setcopyright{none}
 
\begin{document}

\title{IIsy: Practical In-Network Classification}

\author{Changgang Zheng$^{\dagger}$, Zhaoqi Xiong, Thanh T Bui, Siim Kaupmees, Riyad Bensoussane$^{\dagger}$, Antoine Bernabeu$^{\S}$, Shay Vargaftik$^\diamond$, Yaniv Ben-Itzhak$^\diamond$, and Noa Zilberman$^{\dagger}$}
\affiliation{
  \institution{$^{\dagger}$University of Oxford, $^{\S}$École Centrale de Nantes, $^\diamond$VMware Research}
  \city{$\{$changgang.zheng, noa.zilberman$\}$@eng.ox.ac.uk, antoine.bernabeu@eleves.ec-nantes.fr, riyad.bensoussane@worc.ox.ac.uk, $\{$shayv, ybenitzhak$\}$@vmware.com}
}

\renewcommand{\shortauthors}{Zheng, et al.}

\begin{abstract}

The rat race between user-generated data and data-processing systems is currently won by data. 
The increased use of machine learning leads to further increase in processing requirements, while data volume keeps growing. 
To win the race, machine learning needs to be applied to the data as it goes through the network.  \Inl of data can reduce the load on servers, reduce response time and increase scalability.

In this paper, we introduce IIsy, implementing machine learning classification models in a hybrid fashion using off-the-shelf network devices. \name targets three main challenges of in-network classification: (i) mapping classification models to network devices (ii) extracting the required features and (iii) addressing resource and functionality constraints.  
IIsy supports a range of traditional and ensemble machine learning models, scaling independently of the number of stages in a switch pipeline. Moreover, we demonstrate the use of IIsy for hybrid classification, where a small model is implemented on a switch and a large model at the backend, achieving near optimal classification results , while significantly reducing latency and load  on the servers. 

\end{abstract}
\maketitle

\section{Introduction}\label{sec:intro}


Machine learning (ML) is increasingly applied to every aspect of our lives,
leading to huge processing requirements.
In \dcs, \ml has become a prominent workload~\cite{hazelwood2018applied}. To alleviate compute requirements and improve latency-sensitive applications' performance, \ml is pushed to the edge~\cite{cass2019taking} and to end-user devices~\cite{boroumand2018google}.
The performance requirements of \ml have driven the development of a range of \ml accelerators, including GPUs~\cite{awan2017depth}, FPGA~\cite{blott2018finn} and custom ASICs~\cite{jouppi2017datacenter}. While state-of-the-art accelerators can run trillions of operations per second, their throughput is still limited by their network interface. 
Network devices offer an untapped resource for scaling \ml, and in particular -- classification. The use of programmable network devices for \inc applications, such as caching~\cite{jin2017netcache}, consensus~\cite{dang2020p4xos} and network services~\cite{kim2015band}, provides orders of magnitude throughput increase and latency reduction, combined with significant power savings~\cite{tokusashi2019case}. 

\begin{figure}[t]
	\centering
	\includegraphics[width=0.95\columnwidth]{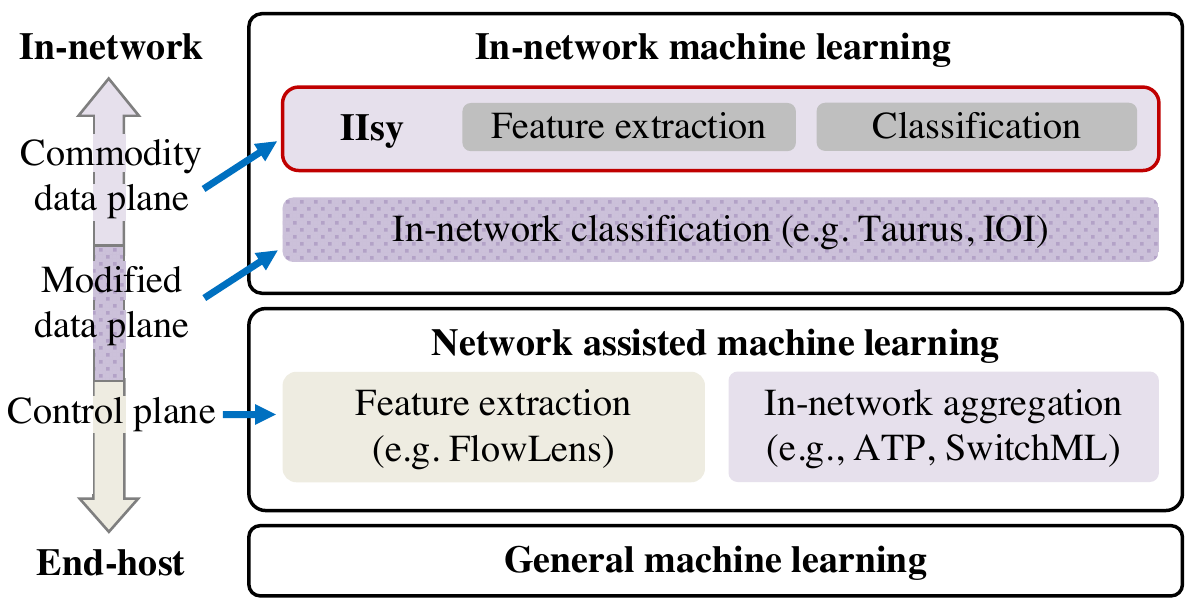}
	\caption{The positioning of IIsy}
	\label{fig:positioning}
	\vspace{-1em}
\end{figure}

Combining \ml and networking is not a new trend (e.g.~\cite{moore2005internet}), with most of the work focusing on \ml on the end host. Newer works had succeeded in creating network-assisted machine learning, as shown in Figure~\ref{fig:positioning}, using network devices either for aggregation~\cite{sapio21switchml,lao21atp}, or for feature extraction~\cite{barradas2021flowlens}. Despite these previous successes, running \ml within network devices has proven hard to tackle. 
While \ml accelerators rely on multiplication and matrix multiplication~\cite{jouppi2017datacenter}, network devices do not support such operations. Several works have tried to attend to this limitation by modifying the data plane or designing new hardware modules~\cite{swamy2022taurus,zhong2021ioi}. 

A few attempts have been made to run \ml models within the network (top of Figure~\ref{fig:positioning}), as detailed in Table~\ref{tab:related_comparison} and \S\ref{sec:related}. The first class of works~\cite{sanvito2018can,siracusano2018network,siracusano2020running,siracusano2022re}, implemented binary neural networks on network interface cards (NICs), FPGA or in a software environment. Their attempts to implement on a switch-ASIC have failed both in scale and performance, as it is significantly more constrained in resources and functionality. The second class of works~\cite{li2019accelerating,swamy2022taurus,zhong2021ioi} enabled \ml by modifying the hardware or using FPGA. These are experimental, not off-the-shelf solutions, and can not be easily and cheaply adopted. The last class of solutions has focused on implementing Random Forests on switches. These solutions had failed to scale on a switch-ASIC~\cite{busse2019pforest} or were independent of most constraints by running in a software environment or on a NIC~\cite{lee2020switchtree, xavier2021programmable}. 
Harnessing the power of network processing for \ml within commodity switches remains a challenge.



To that end, we present IIsy, supporting \textit{off-the-shelf programmable switches} (e.g., Intel Tofino) to employ a range of \ml classification methods. IIsy supports decision-tree, Random-Forest, Isolation-Forest, XGBoost, Support Vector Machine (SVM), Naïve Bayes and K-Means. IIsy is generalizable to other classification methods, but does not support neural network models, to avoid performance compromises. 


The design of IIsy follows the following guidelines: \\
\noindent\textbf{Low-resource ML models} \ml models vary in types and nature, requiring complex mathematical operations, unsupported by switch-ASIC, or consume significant resources (e.g., tree-based models). The scarcity of switch resources means that not every model mapping will be feasible.  Beyond the need to reserve resources for networking functionality, some mappings are impractical, such as using features of hundreds of bytes as lookup keys. Therefore, any model or class of models requires a tailored mapping to a switch architecture. 
We address these challenges in \S\ref{sec:mapping}.

\noindent\textbf{Easy ML model updates} As over time data changes and \ml models need to be re-trained, a quick and easy deployment of \ml models is sought. While switches are programmable, using them in a production environment means that ideally the switch's (P4) program should not be changed, and that only common operations, such as table updates, should be allowed. Moreover, the deployment of an updated classification model should be quick, and minimize traffic disruption. These challenges are addressed in \S\ref{sec:mapping}.

\noindent\textbf{Machine learning performance} For classification purposes, the same level of \ml performance, e.g., accuracy and precision, as running on a CPU or a GPU is targeted.
 While this is highly desirable, it is sometimes practical to trade some accuracy for resources, e.g., saving half the memory resources while giving up 1\% of accuracy. 
 To address this challenge, several possible in-network ML deployments are listed in \S\ref{sec:model}, and \S\ref{sec:evaluation} focuses on the hybrid deployment option, offering competitive \ml performance with low-resource \inl consumption. 
 
\noindent\textbf{Feature extraction} 
While packet-header features can be easily extracted, more complex features are needed to support many \ml models. In PISA-style devices~\cite{bosshart2014p4} this means using the parser to extract specific data from the packet, and the match-action pipeline to turn this extracted data into a feature and store the information. Additionally, it is required to process data stored deep within the payload. These challenges are addressed in \S\ref{sec:feature}.



In summary, this paper presents IIsy -- a framework for automated mapping of trained classification models to commodity, off-the-shelf, switch ASIC. IIsy generates both data plane and control plane programs from the output of a common \ml training framework, and does not require any modifications to tools, network devices, or protocols. 

In particular, our main contributions in this paper are:
\begin{itemize}
    \item Introducing a mapping to programmable network devices of a range of classification methods, including decision-tree, Random-Forest, Isolation-Forest, XGBoost, Support Vector Machine (SVM), Naïve Bayes and K-Means.
    \item Presenting a mapping algorithm that is independent of the number of stages in the switch pipeline, which is critical for scaling ensemble models.
    \item Demonstrating feature extraction on packet, flow, aggregate and file granularity.
    \item Demonstrating IIsy's usability for \inl using a hybrid model, consisting of a small model on a network devices, and large model over the hosts. We demonstrate it can reduce backend's load, and reduce the classification latency for time-sensitive applications.
\end{itemize}

\section{\INL Overview}\label{sec:motivation}


\subsection{Potential Benefits}\label{sec:benefits}

Network devices have two major advantages over any other type of a computing device: location and data-processing speed. Any cloud-processed user-generated data goes through the network first. This means that i) the rate of data that can be processed is capped by the network, ii) the latency from the user to the processing node will always be higher than the latency to any network device along its path and iii) network devices are already part of the infrastructure carrying user data, and do not need to be newly added. This leads to the observation that \textit{the rate of classification decisions is bounded by network devices' data rate}.

From a system perspective, \inc was already shown to be beneficial, freeing cycles on the CPU and providing high power efficiency per operation~\cite{tokusashi2019case}. Improved throughput and latency are also known advantages~\cite{jin2017netcache,dang2020p4xos}, but need to be considered per use-case. As we show later, \inl can significantly reduce the amount of traffic that gets to servers, and that requires further processing.

Using network devices for in-network \ml suggests a few more benefits. Firstly, being able to classify data before it reaches the host can be essential for some use-cases. For example, distributed denial of service (DDoS) mitigation, where malicious traffic has to be dropped as close as possible to the source. Secondly, automatically converting and loading \ml training results to (local and remote) network devices~\cite{iisy}, can speed up the reaction to events in the network, and shorten the time for detection and mitigation.

\subsection{Deployments Scenarios}\label{sec:model}

\Inl is possible in different deployment scenarios, including: (1) a native switch operation, (2) a switch acting as an endpoint accelerator, (3) smart NICs, and (4) a hybrid \ml model, combining the \inl model with a traditional \ml model deployed at the end-point.

\textbf{Native switch.}
A switch in its native usage model has some, or most, of its resources dedicated to networking operations, meaning that \inl needs to be resource efficient. Using \inl within the switch does not require extra cost or space, and the power overheads are small~\cite{tokusashi2019case}. The location of a switch has affects its benefits; A switch very close to the user is most useful for data reduction, ultra low latency applications, and to mitigate the effects of distributed events (e.g., DDoS attacks). On the other hand, a switch within a \dc can support more complex applications. For example, assuming that the switch is located after a load balancer, decrypted traffic can sometimes be assumed~\cite{aws_lb,nginx_lb}, allowing to apply \inl to use-cases otherwise prohibited by traffic encryption.

\textbf{Endpoint accelerator.}
The switch as an endpoint accelerator refers to using a switch purely for \ml purposes. This model is already in use for some applications, such as load balancing~\cite{insidepacket2020lb}. 
Unlike other deployment scenarios, here the switch adds space, power, and cost overheads.

\textbf{Smart NIC.}
Smart NICs support lower bandwidth than switches but benefit from better resources availability, such as on-board memory and encryption modules. The host's proximity means that the devices see only a subset of the overall network traffic. Host offloading and power savings are reduced relative to a switch~\cite{tokusashi2019case}. Our solution is applicable to architectures like the Portable NIC Architecture (PNA)~\cite{brebner2018extending}.

\textbf{Hybrid. }
The resource-constrained nature of network devices means that in-network ensemble models are smaller than full-grown ensemble models, hence their \ml performance may be sub-par. In such cases, a \textit{hybrid \ml model} can be used to achieve close to optimum \ml performance, while still benefiting from the performance of \inl. The hybrid deployment scenario employs a small in-network \ml model and a large \ml model over the end-point.

\subsubsection{Hybrid Deployments}\label{sec:hybrid}
In many \ml ensemble models, such as Random-Forest and XGBoost, the \ml model can provide a classification with a corresponding confidence level --  the probability that the classification is correct.

In this paper, we adopt the hybrid \ml model concept~\cite{silla2011survey,vargaftik2019rade}, by implementing a small in-network ensemble model (e.g., by limiting the number of trees in an ensemble, or using a subset of features~\cite{tang2014feature}), and running a large \ml model at the end-point.  
To cope with the lower \ml performance of the small in-network \ml model, classifications by the small model are considered valid only if their corresponding confidence is above a given (high) threshold. Invalid classifications by the small in-network \ml model (i.e., confidence below the threshold) are forwarded for re-classification by the large \ml model deployed at the back-end. 

Previous \ml works~\cite{silla2011survey,vargaftik2019rade} have shown that most of the queries in a given data-set can be classified by a small \ml model with high confidence level. Hence, a hybrid \ml deployment reduces both the classification latency and the load over the back-end servers (by forwarding only ``hard'' queries for re-classifications), as compared to a monolithic \ml model deployment at the back-end.
In \S\ref{sec:evaluation}, we demonstrate these benefits using two use-cases from different domains, cyber-security and finance.

\section{IIsy Architecture}\label{sec:arch}

 IIsy is a framework that automatically maps trained classification models to programmable network devices, and in particular to off-the-shelf switch ASIC. IIsy takes the output of a common \ml training framework, and converts it to data plane code and control plane code. 

\begin{figure}[htb]
	\centering
	\includegraphics[width=1\columnwidth]{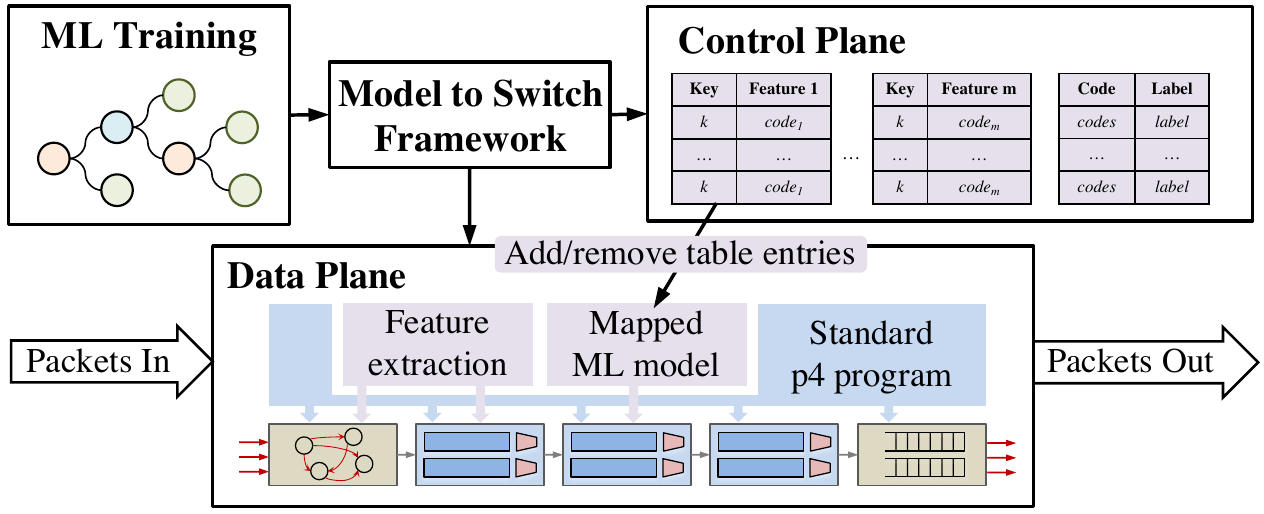}
	\caption{The architecture of IIsy}
	\label{fig:arch}
\end{figure}

The architecture of \name, shown in Figure~\ref{fig:arch}, is composed of four components: machine learning training, mapping tool to map the trained models to a target network device (the core of \name's architecture), a data plane implementation on a hardware target, and a control plane component for populating table entries. 

\name's mapping tool takes a standard \ml training output (e.g., pickle file) and generates from it two components: an implementation of the data plane, and the entries for the \lut used by the data plane which are loaded by the control plane. 
The hardware target and the control plane component contain standard elements that are target specific, such as the architecture of the network device (e.g., Tofino Native Architecture, TNA). 
In the following section, we present \name's mapping tool in details.

\section{Mapping Models to Switches}\label{sec:mapping}

IIsy's mapping of \ml classification model to network devices is guided by several insights~\cite{iisy}:
\begin{itemize}
    \item Use \lut to implement mathematical operations, for example multiplication and exponents.
    \item Optimize the use of on-chip resources, and reduce \lut size by using codes or indicators instead of  explicit calculation results.
    \item Use a \lut to decide a classification result at the end of the pipeline. The key to this table is composed of a set of indicators collected from the previous stages.
    \item Be willing to lose some accuracy to save resources. Decide on the (accuracy) price to pay to fit a model into your network device, or use a hybrid deployment.
    \item Break the dependency between tree-depth and pipeline-stages by looking up values instead of conditions.
    \item  Optimize the use of on-chip resources by sharing features \lut between multiple trees or models.
\end{itemize}

In the following, we present our mapping method for different \ml models to a network device. 
We discuss the mapping challenges involved with each \ml model and the corresponding mapping solution.

\subsection{Decision Tree}\label{sec:DT}
Decision-trees naturally fit into network devices, since their most basic functionality is packet classification --  every ingress packet is assigned to an output port (i.e., class in terms of \ml). To that end, \lut are used for classifying packets to their corresponding output port. For instance, in a layer-2 Ethernet switch, the feature used for classification is the destination MAC address, and the MAC table is used to decide the output port -- the classification's result. 

\name uses a more efficient mapping of a decision tree to a network device by using a single table per feature\footnote{Features are extracted as in \S\ref{sec:feature}}, and a single classification table.

 \begin{figure*}[t]
	\centering
	\includegraphics[width=0.85\linewidth]{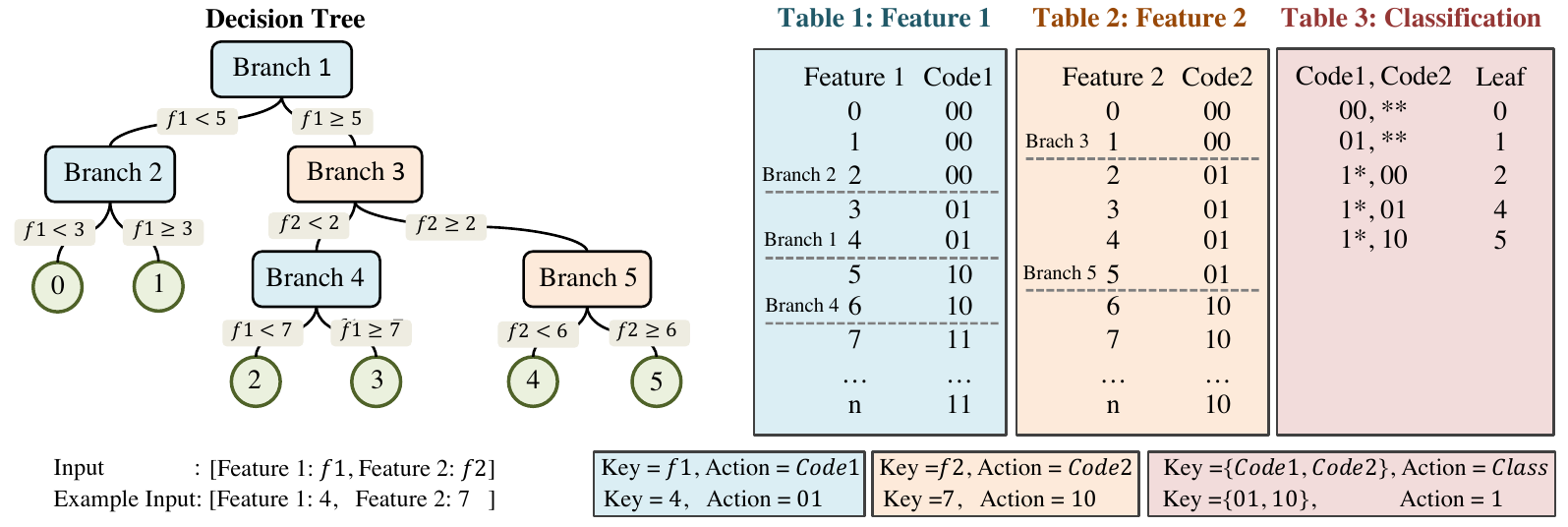}
	\vspace{-1em}
	\caption{An Example of a Decision Tree Mapping.
}
	\label{fig:model_dt_toy_exp}
 	\vspace{-1em}
\end{figure*}

\name's mapping is demonstrated using a simple example in Figure~\ref{fig:model_dt_toy_exp}. The example shows a decision tree with a depth of three,  with six leaf nodes, and using two features. The color of each branch in the tree indicates the feature used, with three branches using feature 1 ($f1$), and two branches using feature 2 ($f2$). The branches using feature 1 are mapped to Table 1, with  feature 1 used as the key to the table. The table has 4 ranges, covering the potential outcomes of branches 1,2 and 4.
Each range is associated with a code as the resulting action. Similarly, Table 2 uses feature 2 as the key, with 3 ranges,
 corresponding to branches 3 and 5. The action of this table is a second 2-bit code. While the tables are shown as exact match, a ternary implementation is possible. The third table, the classification table, uses as the key the codes (actions) of Table 1 and Table 2. A match on this key results in a leaf node, the result of the classification. 

Each leaf node is denoted by a code word, and the code word indicates the branches taken in the tree. This code word is the result of the feature tables lookup. Therefore the number of pipeline stages consumed by a decision tree is independent of the depth of the tree, overcoming previous limitations~\cite{busse2019pforest,lee2020switchtree}.

\textbf{Table size analysis.} Assume a tree with $B$ branches using $F$ features, where each feature $f_i$ is $w_i$ bits wide and used in $b_i$ branches. The number of entries in a ternary feature table for $f_i$ will be $O(b_i\times w_i)$, while an exact match table will contain all the feature's values that are possible for the use case. This number can end up small, as shown in \S\ref{sec:eval-resources}, for example if most values are mapped to a default entry.

The number of entries in the classification table depends on the depth and shape of the tree. The worst case is when the branches are evenly divided between all features, so features require the same number of (multiple) bits for the code. The best case is where $F-1$ features are used only once in branches, requiring a single bit code, and the last feature is used $B-(F-1)$ times. This can be written as:
\begin{equation}
2^{  \lceil\log_2 \frac{B}{F}\rceil^F} \geq Entries \geq 2^{B-1+\lceil\log_2(B-(F-1))\rceil}
\label{eq:code_bound}
\end{equation}

\subsection{Ensemble Tree-Based Methods}\label{sec:ensemble}

Ensemble methods improve \ml prediction results by combining multiple learning models~\cite{zhou2012ensemble}. We consider two types of ensemble methods: Bagging and Boosting.

The size of an ensemble is known to improve a prediction's accuracy, which stands in contradiction to the scarcity of resources on network devices. To attend to this challenge, we maintain the guidelines presented in \S\ref{sec:mapping}, such as using a single table per feature and combining multiple models within a single table. In this manner, we increase tables' depth and the amount of metadata used but decouple the number of trees from the number of stages required in the device.

\textbf{Bagging.}\label{sec:RF}
In bagging, multiple learners are used, and each learner has an equal weight in the final decision. Each learner is trained using a different sample with replacement of the training data. We use Random Forest~\cite{breiman2001random}, which is built from multiple decision trees, as an example of mapping a bagging model to a network device.

The mapping of a model to a switch is oblivious to differences between bagging models in terms of training related parameters such as sample selection and training method. The mapping is influenced only by the constraints of the training outcome, e.g., selected features, number and depth of trees.

While each decision tree in a Random Forest can be independently mapped to a device, as in \S\ref{sec:DT}, this is inefficient. For example, a Random Forest of ten trees, each using five features, will require fifty feature tables and eleven decision actions (one per tree, plus one for the entire forest).

IIsy significantly reduces resources requirements by sharing feature tables between trees. This means that for the previous example, \name will require just five feature tables instead of fifty. 
This mapping is not free; the number of entries in each table will increase, as well as the action's width. The result of each lookup in a feature's table is the series of action codes (as defined in \S\ref{sec:DT}), one per tree. Trees can be pruned to create action codes of feasible length. 
As demonstrated for a single tree, \name requires a table per tree to turn the action code into the decision of a tree. The classification result of the entire ensemble is based on the collective classification results of all trees in the ensemble, and can be implemented as a table or using a sum and conditions.

\textbf{Boosting.}\label{sec:boosting}
Boosting methods are different from bagging \cite{schapire2013boosting}, as the ensemble is built by training new learners to focus on misclassifications by previous learners.
Gradient boosting is often built from an ensemble of decision trees, where a small decision tree (e.g., with 8-32 terminal nodes) is added at each iteration and scaled by a constant factor. Then, a new tree is grown to reduce the loss function of the previous trees. In boosting, new trees are trained with a focus on previous misclassifications. The decision is based on the weighted outcome of each tree. 

Despite the differences in training, the mapping of a generated XGBoost model is mostly identical to a Random Forest (\S\ref{sec:RF}), using a table per feature, and a table per tree. The difference is that the leaf nodes are weighted. The weighting can be applied either when constructing the tree table, which is typically more resource-efficient, or at the decision stage. When the number of classes is small, it is effective (\S\ref{sec:evaluation}) to add the weighted results across trees, either directly or using a \lut, and setting the class at the final stage. 
Isolation Forests are implemented similarly, but summing the depth of terminal nodes, rather than their weights.

\subsection{Classical models: SVM, Naïve Bayes and K-Means}\label{sec:map-general}

Classical classification algorithms can all be mapped using similar methodologies. These are applied to SVM, Na\"{\i}ve Bayes and K-Means.
A mapping takes one of two forms. The first holds one table per feature. The result of looking up the feature in the table is a code, or a value that is normalized. If the result of a lookup is a code, the last stage in the pipeline will use a \lut with a key of the codes of all features. If the result of a lookup is a value, then the last stage in the pipeline will operate on all values, typically adding them up and comparing the results across classes. 

For example, in SVM, the key to a feature's table is the feature's value, and lookup's result is a vector of calculated values $a_{i}\times x_{i}$, where $x_{i}$ is the value of the feature (potentially normalized or binned). The value of an SVM hyperplane, separating two classes, is calculated as the sum of vectors from all feature tables. This can be optimized by summing the features in each pipeline stage. 

A second approach, which is not always feasible, holds a table per class or class indicator. For example, in SVM there will be a table per hyperplane. The lookup key is the value all the features. The result of the lookup will be an indicator, such as if the entry belongs within or outside the hyperplane (for SVM) or the distance from a center of a cluster (for K-Means). 

Our experience shows that the first approach provides (relatively) shallow tables, proportional to the number of classes. However, this approach may experience some loss of accuracy, explored in \S\ref{sec:eval-ml}. The second approach is feasible only when the use of multiple features leads to a feasible key size and table depth. It provides higher accuracy results and requires fewer operations at the last stage. 

Clearly, no single solution fits all use-cases. The approach fits some classification, regression, and clustering models. Iterative models are less suitable, though they may be feasible, e.g., using recirculation. Importantly, the framework takes care that the type of feature or its range will not affect the accuracy of the classification (e.g., through normalization).

\subsection{Retraining and Updates}

\ml models often need to be retrained and the resulting classification model needs to be updated. IIsy enables updating a model deployed on a switch using only table updates, without changes to the deployed program.

For a given use case, a user defines the features that need to be extracted, the type of the model, and, in the case of an ensemble model, the constraints on the model (e.g., number of trees). This leads to a generated P4 program for the network device. As long as the user maintains the constraints, retraining the model will not change the P4 program. Still, retraining will result in a different \ml model, mapping to different actions in the features table, and in different code-to-classification entries in the tree and decision tables (as in Figure~\ref{fig:model_dt_toy_exp}). These can be updated by table updates, a common management operation. 

In a hybrid deployment, traffic can be directed to the back-end during updates, to avoid misclassification.

\section{Feature Extraction}\label{sec:feature}
Network devices are designed to extract headers from packets. However, the research community has already gone beyond packet headers for applications ranging from telemetry~\cite{kim2015band} to \inc~\cite{tokusashi2019case}. In this section, we discuss how features can be extracted from data on different levels of granularity.
While we use PISA and P4-nomenclature\cite{bosshart2014p4}, similar concepts apply to other targets. 

\subsection{Packet \& flow level features}
Extracting packet-level features is native to network devices, with packet header extraction done in the parser, and features are stateless. Such features include, for example, protocol type or source and destination port number.
Packet level features also refer to features that describe the packet, such as packet size, switch source port, or timestamp. 

Flow level features are stateful, and information is collected and stored across multiple packets. Examples of flow-level features include flow size, flow duration, and flow data rate. 
 Heavy hitters detection is one line of research where flow-level features are already extracted and used~\cite{sivaraman2017heavy}. 

We distinguish between two types of flow-level features: counted features (e.g., flow size, packets count), and time-related features (e.g., flow's start time, inter-packet gap).

\subsection{Aggregate level features}
Aggregate level features consider a group of flows,
an aggregation of traffic (e.g., traffic from/to port $X$) or the network as a whole. Examples of features useful for \ml purposes include traffic volume from a group of subnets, inter-arrival time toward a specific application or a histogram of source and destination ports. 
Aggregate level features are mostly similar in implementation to flow level features, however they may require additional operations, such as mapping flow-identifiers to an aggregated-feature identifier.

\subsection{File level features}

Extracting features from a file is more complex than any previous case, yet building upon flow-level feature extraction makes the challenge simpler. 
We distinguish between four stages of file processing: 

\noindent$\bullet$ Start of a file, where file header needs to be processed, and initial resources need to be assigned. This is similar to a start of a flow but with a more complex parsing of the header.\\
\noindent$\bullet$  Looking into the file's payload. If a packet exceeds the size of the programmable data plane bus, then it may need to be recirculated (target dependent).\\
\noindent$\bullet$  Examining payload across packets. As a file is likely to be broken across many packets, extracting features from a file means that contents at the end of a previous packet need to be stitched with the contents at the head of the next packet. \\
\noindent$\bullet$ End of file. This is similar to the end of a flow and may allow to free up some resources.

To be clear, it is feasible to extract data from a subset of filetypes, not from all file types. Text-based files, such as txt, xml, html and csv, are straightforward to process. File types that use many objects, such as docx, pdf and xlsx, are very hard to process due to the complex structure and required resources. Certain image file types, e.g., png and tif, or audio files such as mp3, have a complex file structure and require significant switch resources, making them impractical. Video files, composed of frames of images, will be even harder to process. One exception to image file types is JPEG, which is relatively easier to process.  Extracting a feature from the JPEG file, such as the average value or the value of a certain pixel, is possible. However, extracting more complex features will likely be beyond the resource budget~\cite{glebke2019towards}.

Handling files raises other concerns. First, we assume that files are not encrypted (\S\ref{sec:model}). Second, privacy and legal rights to process the data need to be addressed by the operator. Third, we assume no packet reordering, e.g., direct-attached SmartNIC or endpoint accelerator. Last, we assume file-type specific feature extraction. 

We focus on two of the challenges raised by file processing: looking within the packet's payload and examining payload across packets. To look deep into the packet, it needs to be recirculated, with bytes that were already processed stripped from the packet. This process is destructive, as the removed data can not be returned when the processing is done. Furthermore, recirculation can lead to significant bandwidth loss. 

Examining payload across packets\footnote{For recirculated packets one can strip data on feature boundaries.} creates resource challenges. Assume that you want to access a feature partly stored in the last two bytes of a packet and partly in the first 2 payload bytes of the next packet. You need to save the last 2 bytes of the packet in the memory (e.g., register) and the amount of data that was saved (i.e. two bytes). However, this information can be accessed only in a single place in the pipeline. If the last two bytes were saved at a stage toward the end of the pipeline, it would not be possible to extract this data until the new packet reaches this stage. One solution is recirculation, where the first pass through the pipeline extracts the missing information, potentially adding it as metadata or a header. Another solution is to extract this information in the Ingress pipeline and process it in the Egress pipeline, but this limits the program's functionality. 

There is a trade-off in functionality, performance, and resource efficiency when applied to specific file-level use cases. As previous works have suggested, an easy get away is to manipulate the file sent at the host's side before entering the network so that some challenges can be avoided.

\section{Implementation}\label{sec:implement}

IIsy's framework uses four components, as described in \S\ref{sec:arch}. In this section, we describe the implementation of our prototype. Further generalization of IIsy's framework is described in~\cite{zheng2021planter,planterarxiv}.

The prototype's machine learning training framework is based on \textit{scikit-learn}~\cite{scikit-learn}. Our implementation enables fast development and prototyping of different models and, in particular, the hybrid approach. 
The training of the hybrid models used scikit-learn 0.24.1 and XGBoost 1.3.3, running over a \texttt{c4.8xlarge} AWS EC2 instance with 36 vCPUs and 60~GB RAM running Ubuntu 16.04 LTS.

The switch implementation run on two platforms: Intel's Barefoot Tofino (ASIC), and NetFPGA-SUME~\cite{zilberman2014sume}(FPGA). All the models are mapped to both targets, except for boosting, which targets only Tofino. The NetFPGA implementation enables exploring the limits of feature extraction. 
This includes also complex stateful features~(\S\ref{sec:feature}), such as jitter, inter-arrival time, and data rate. On Tofino, packet-level, flow and aggregate features are supported, with further focus on files. Data is extracted from text files, both where the size of a feature is known (constant) and for unknown feature length (e.g., words separated by delimiters). Our implementation currently supports features of up to 15 ASCII characters per feature. In addition, it supports features split between packets and features implemented deep within the packet (\S\ref{eval:finance}).

The simplicity of the mapping enables to auto-generate the data-plane and the control-plane, using a python script and a configuration file. A user defines in a configuration file design constraints, such as maximum number of trees, and the tool takes the output of the training stage (pickle file) and uses it to generate both the data plane (P4 files) and the control plane (table entries, in json format).

The system test environment 
uses $64\times100G$ ports Barefoot Tofino.  P4-NetFPGA~\cite{ibanez2019p4netfpga} is used for FPGA development. Four servers with 100G NICs are used to send and receive traffic from the switch. To test full throughput, we use a snake configuration, where traffic is looped from each port to the following one, enabling traffic across all 64 ports, which is a common practice~\cite{dang2020p4xos}. As a baseline, we measure 6.2Tbps on the switch when running simple forwarding.

\section{Evaluation}\label{sec:evaluation}

In this section, we evaluate \inl for feasibility, performance, resource consumption and \ml performance.
For brevity, this section focuses on Intel Tofino, and details of the NetFPGA evaluation are provided in 
~\cite{iisyrepo}.

\subsection{Use cases}

Our evaluation is driven by two use cases: network anomaly detection using the UNSW-NB15 dataset~\cite{moustafa2015unsw}, and time sensitive financial transactions using the Jane Street Market Prediction~\cite{janekaggle}. For each of these use cases, described below, we explore the classification performance of the switch alone, as well as part of a hybrid model\ref{sec:ensemble}.

\subsubsection{Anomaly detection - Reducing back-end resource consumption}\label{sec:eval-unsw}

Anomaly detection, such as intrusion detection and prevention, is typically done at the back-end and can consume significant compute or acceleration resources~\cite{zhao2020achieving}. All network traffic toward certain application servers needs to be examined, and non-human or malicious traffic needs to be filtered. Our goal is to provide a scalable solution, whereby normal traffic is allowed in by the switch, and anomaly traffic is either dropped (where all decisions are taken by the switch) or sent to the back-end (in a hybrid mode). The aim of the system is to allow in all normal traffic, which is the majority of traffic volume. In the hybrid mode, traffic that is classified as anomalous or with low confidence is sent to the backend for deeper inspection. In this manner, the switch does not block (drop) legitimate traffic and offloads significant processing from the backend, as most traffic is normal. 
This use-case is an example where \inl saves resources compared with host-based solutions while also scaling with the network's bandwidth.

To demonstrate this use-case, we use the UNSW-NB15 dataset~\cite{moustafa2015unsw} that contains a mix of normal traffic and different types of attacks. 
The goal of the prediction is to detect attack traffic, which we label ``anomaly'' in our evaluation.

The use case is explored for the various \ml models (\S\ref{sec:mapping}), for feasibility study purposes. From \ml perspective, Random Forest is the most suitable for this use-case, as it offers low variance in its classifications. This leads to a more predictable fraction of the traffic that is correctly classified as normal (unless the traffic distribution changes dramatically -- which requires retraining the model).

Our learning uses 80\% of the data for training and 20\% for testing. The model running on the back-end is using a Random Forest of 200 trees (estimators) and 10,000 leaf nodes, and all the features in the dataset.

\subsubsection{Financial transactions - Reducing latency}\label{eval:finance}

Low latency financial transactions, such as algorithmic trading, are very sensitive to latency. The lower the latency of a transaction, the higher the potential gain, even with slight reduction in latency.
For top 10\% financial traders, the a decision latency is less than 42 microseconds~\cite{baron2019risk} from a passive order to an active transaction. 

Typically, a large backend is used to provide real time classification for all transactions. 
In this case, the switch can be used to identify and tag high priority transactions, while other transactions are sent to the backend for fine-grain classification. The tagged high priority transactions can be forwarded to a different for immediate execution. Moreover, tagged queries can be prioritized over a dedicated link(s), avoiding congestion. 
While the switch may miss some high-priority transactions, those, in turn, will undergo the regular classification path. The gain in assigning many of the time-sensitive requests to a special fast processing path may introduce significant financial benefits with low resource consumption. This is an example where the advantage is the latency of classifying high-priority events, while the change in the backend's load is small. 

To demonstrate this use-case, we use the Jane Street Market Prediction dataset \cite{janekaggle}, a recent trading finance dataset. Each trade in this dataset contains 130 anonymized features representing real stock market data and two output values ('weight' and 'resp') representing the trade's return. According to these two output values, we label the transactions in the dataset by recommended actions: 'Strong sell or buy', and 'Sell/Hold/Buy'.
Financial transactions are typically a feed of individual trade instructions, arriving from the stock exchange. Such a feed is not openly available. The Jane Street dataset is the most recent and open information available from a trading company, presenting pre-processed transactions. 

Our goal is to minimize the latency experienced by transactions marked as ``strong or sell'' (accounts for $\approx$13.1\% of the total transactions). We assume the switch is located in such point in the network that any incoming transaction must go through it, so any classification by the switch has an additive latency of close to zero\footnote{Some settings use L1 switches, such as Cisco Nexus 3550. This is a different use case.}.

In terms of \ml performance, while we evaluate with different models, the target for this use case is XGBoost, commonly used in financial applications as boosting offers a controlled bias which is more suitable for identifying minority.

Our learning uses 80\% of the dataset for training and 20\% for testing. The model running on the back-end is using all 130 features, with XGBoost of 100 trees (estimators) and a maximum depth of 8 (XGBoost trees tend to be shallow).

\subsection{Feature Extraction}

In the anomaly detection use-case, we implement on Tofino support for packet level features (e.g., source and destination port, protocol, service, and ports equivalence) and flow level features (e.g., duration, flow size in bytes and packets in each direction). 
While flow level features can improve the quality of the prediction, they cost two stages within the switch: to hash the flow ID, and to update a register holding the features value (e.g. flow size). Choosing between the two options requires weighting also other considerations, such as if flow ID is needed for ``standard'' processing purposes. Our resource consumption evaluation uses the features 'sport', 'dsport', 'proto', 'service', and 'is\_sm \_ips\_ports' (Table~\ref{tab:unsw_resource_performance}), and the study of ensemble models (Table~\ref{tab:ensemble_scalability}) uses the features 'sport', 'dsport', 'proto', 'service', 'sbytes'. This is as the 'sbytes' feature only improves the performance of the ensemble models.

The Jane Street dataset contains 130 numerical features, which we process two ways: either as a packet containing the features as numerical values, or in its original csv format, demonstrating the feasibility of file processing. 
For ease of exploration we reformat the file as columns of eight characters, but note that other implementations under this work are not of fixed size or known delimiter location. Both numerical and csv formats allow to explore feature extraction from deep within the packet. We succeed in extracting features from any of the 130 columns, without recirculation. The limitation to the extraction is not the location of the extracted feature, but rather the maximum number of features extracted and their size, which are limited by parser's resources.
As financial transactions are typically a feed of individual trade instructions (\S\ref{eval:finance}), and the size of an entry in the Jane street dataset, with 130 columns, barely fits within an MTU packet (1522B), we send each transaction as a separate packet. 
The features that we use are features 42, 43, 45, 124 and 126, demonstrating our ability to look deep into the data. 

\subsection{Resource Consumption}\label{sec:eval-resources}

We run an exploratory experiment where the goal is to maximize the performance of the \ml prediction, while still fitting the design within the ingress pipeline. SVM, Na\"{\i}ve Bayes, K-Means and Decision Tree (DT) fit within the ingress pipeline. While the Egress pipeline can be used, it is not recommended, as discussed in \S\ref{sec:discussion}.

Table~\ref{tab:unsw_resource_performance} and Table~\ref{tab:finance_resource_performance} summarize the resource consumption of the anomaly detection and financial transactions implementations, respectively. The tables show, for each model, the maximum size of the model that fits within an ingress pipeline switch using 5 features. The memory and latency are measured in comparison with Tofino's switch.p4 reference design.
The number of tables does not directly map to the number of stages, and can be significantly higher. Our p4 programs see multiple feature tables mapped to the same stage.

As the results show, the memory requirements (all SRAM) are quite low in comparison with switch.p4, despite their potential to scale. This demonstrates the efficiency of our mapping algorithm. For Bayes, the results are relatively high as we maximize the accuracy in our features tables, and as we use two table to calculate the final probability (multiplication). In contrast, SVM and K-Means can have their results added up, without a decision, which saves significant resources.

\begin{table}[]
\begin{adjustbox}{width=\columnwidth,center}
    \centering
    \begin{tabular}{|l|c|c|c|c|c|c|}
    \hline
      Model  & SVM &  Bayes & KMeans & DT & RF & XGB\\
         \hline
         Tables & 5 & 7 & 5 & 6 & 10 & 10 \\
         \hline
         Memory & 5.7\% & 15.7\% & 6.0\% & 2.5\%& 9.4\%&  3.6\%\\
         \hline
         Stages &    5   &  6     & 6  &    3  &  5    &  4       \\
         \hline
         Latency & 34.2\% & 40.8\% & 34.6\% &33.6\%& 45.9\%&  39.4\%\\
         \hline
         F1 &  0.815 & 0.846 & 0.815 & 0.848 & 0.848 &  0.849 \\
        \hline
    \end{tabular}
  \end{adjustbox}
\vspace{1em}
\caption{Anomaly Detection - Resource consumption of the models on Intel Tofino, using 5 features. 2 Stages are required for feature extraction and sending the packet. Resource Consumption is relative to switch.p4 reference program. }
    \label{tab:unsw_resource_performance}
   \vspace{-1em}
\end{table}

\begin{table}[]
    \centering
    \begin{adjustbox}{width=\columnwidth,center}
    \begin{tabular}{|l|c|c|c|c|c|c|}
    \hline
      Model  & SVM &  Bayes & KMeans & DT & RF & XGB\\
         \hline
         Tables & 5 & 7 & 5 &  6 & 15 & 9 \\
         \hline
         Memory & 1.8\% & 14.7\%  & 1.7\% & 12.1\% & 5.5 \% & 3.9\% \\
         \hline
         Stages &    5   &  6     &    6    &    3  &   3  &    3     \\
		\hline
         Latency & 27.4\% & 33.9\% & 27.7\% & 32.9\% & 32.2\% & 32.2\% \\
         \hline
         F1 & 0.639 & 0.696 & 0.705 & 0.678 &  0.884 & 0.677 \\
         \hline
    \end{tabular}
    \end{adjustbox}
    \vspace{1em}
    \caption{Financial Transactions - Resource consumption of the models on Intel Tofino, using 5 features. 1 stage is required for sending the packet. Resource Consumption is relative to switch.p4 reference program.}
    \label{tab:finance_resource_performance}
   \vspace{-2em}
\end{table}

\subsection{Scalability}\label{sec:scalability}

\begin{figure}[t]
	\centering
	\begin{minipage}{0.49\linewidth}
	\includegraphics[width=1\linewidth]{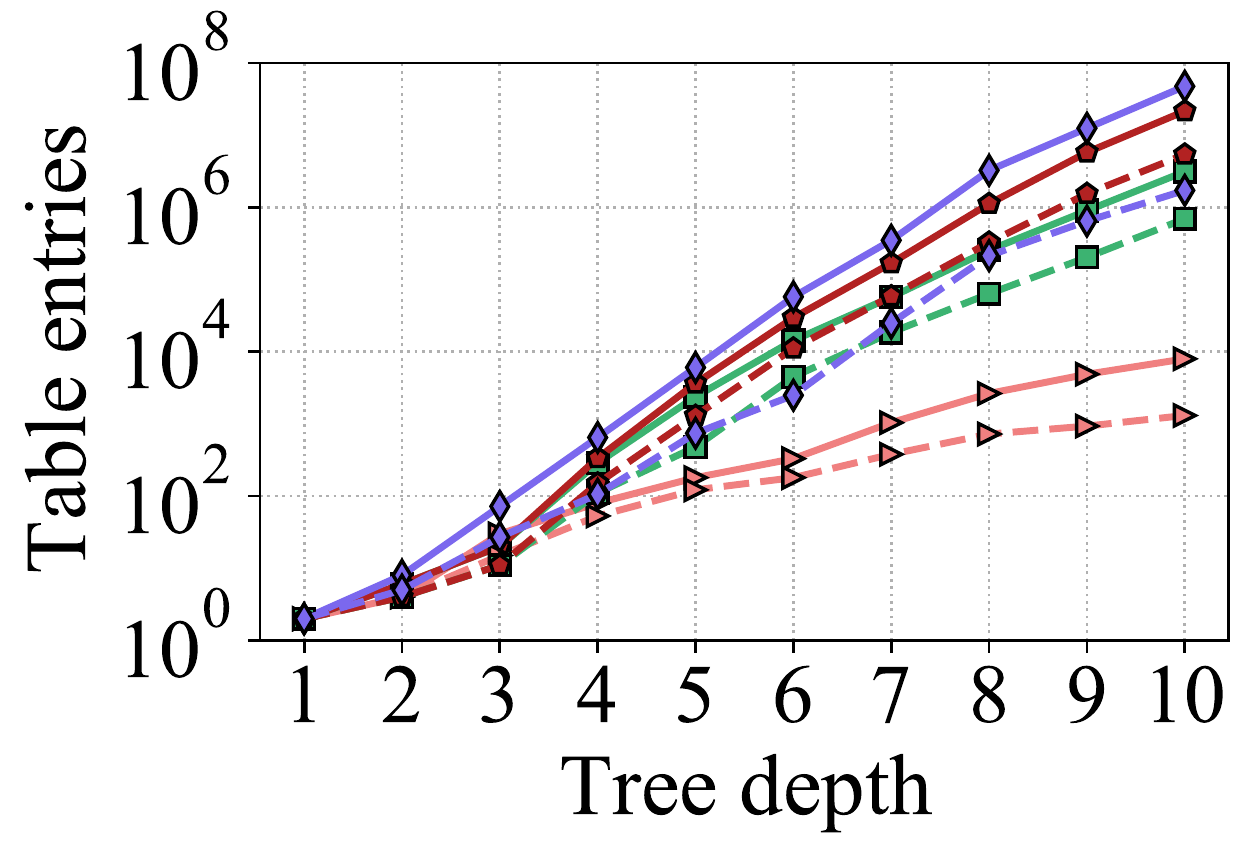}\vspace{-0.5em}\\\centering(a) Anomaly Detection
\end{minipage}
	\begin{minipage}{0.49\linewidth}
	\includegraphics[width=1\linewidth]{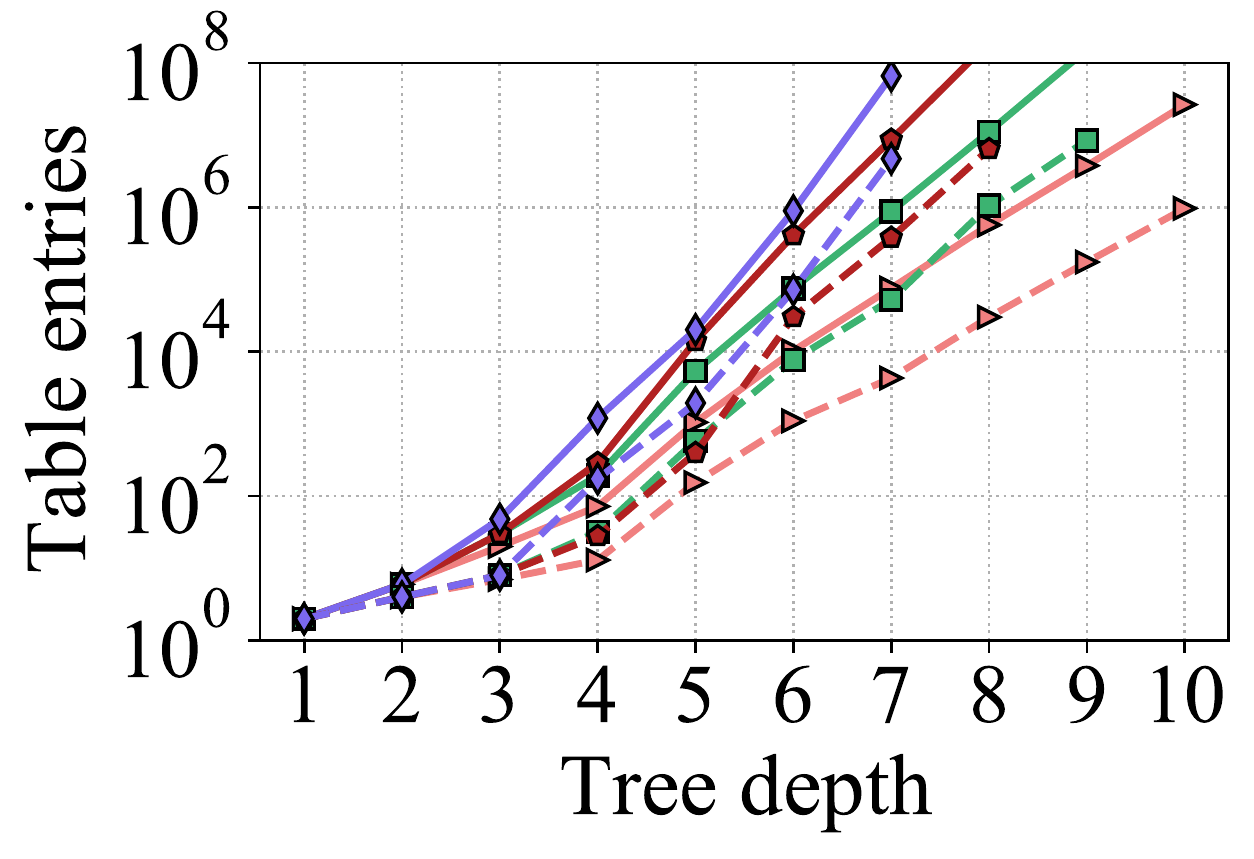}\vspace{-0.5em}\\\centering(b) Financial Transactions.
\end{minipage}
\vspace{0.5em}

	\begin{minipage}{0.49\linewidth}
	\includegraphics[width=1\linewidth]{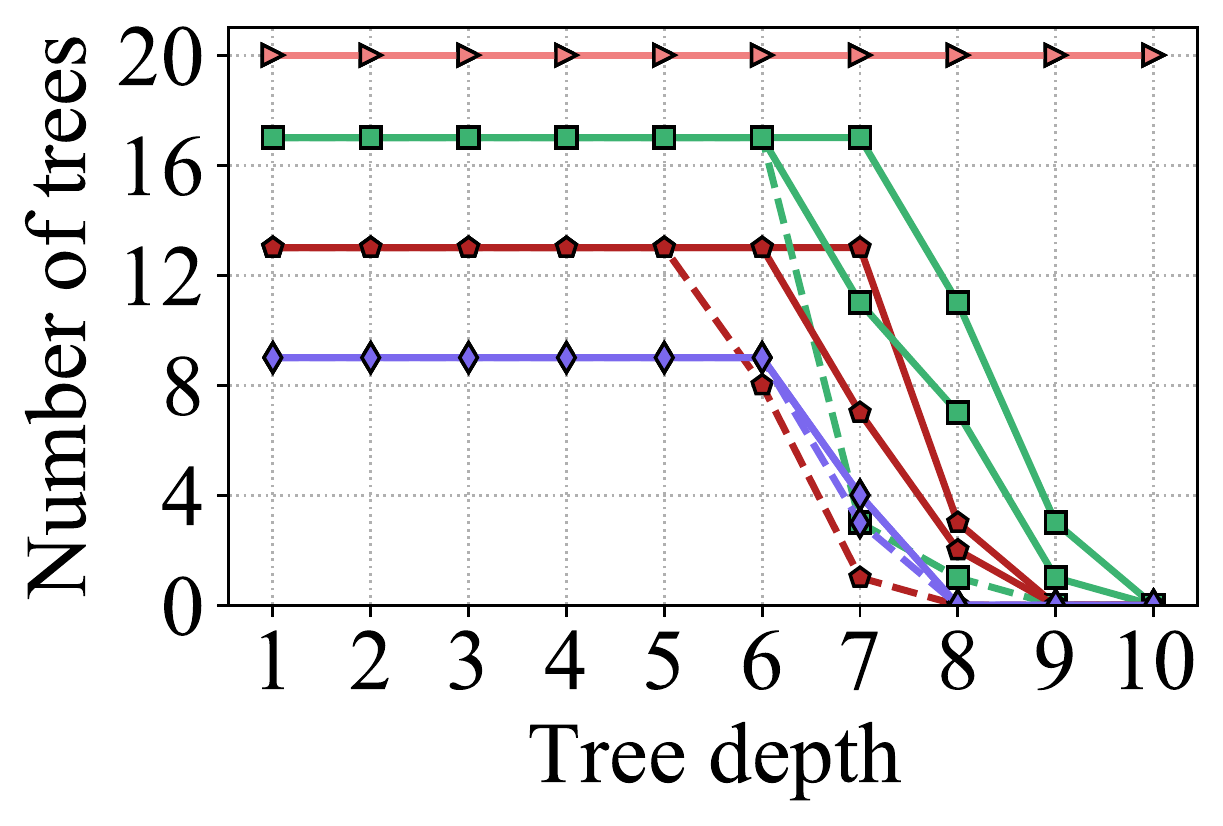}\vspace{-0.5em}\\\centering(c) Anomaly Detection
\end{minipage}
	\begin{minipage}{0.49\linewidth}
	\includegraphics[width=1\linewidth]{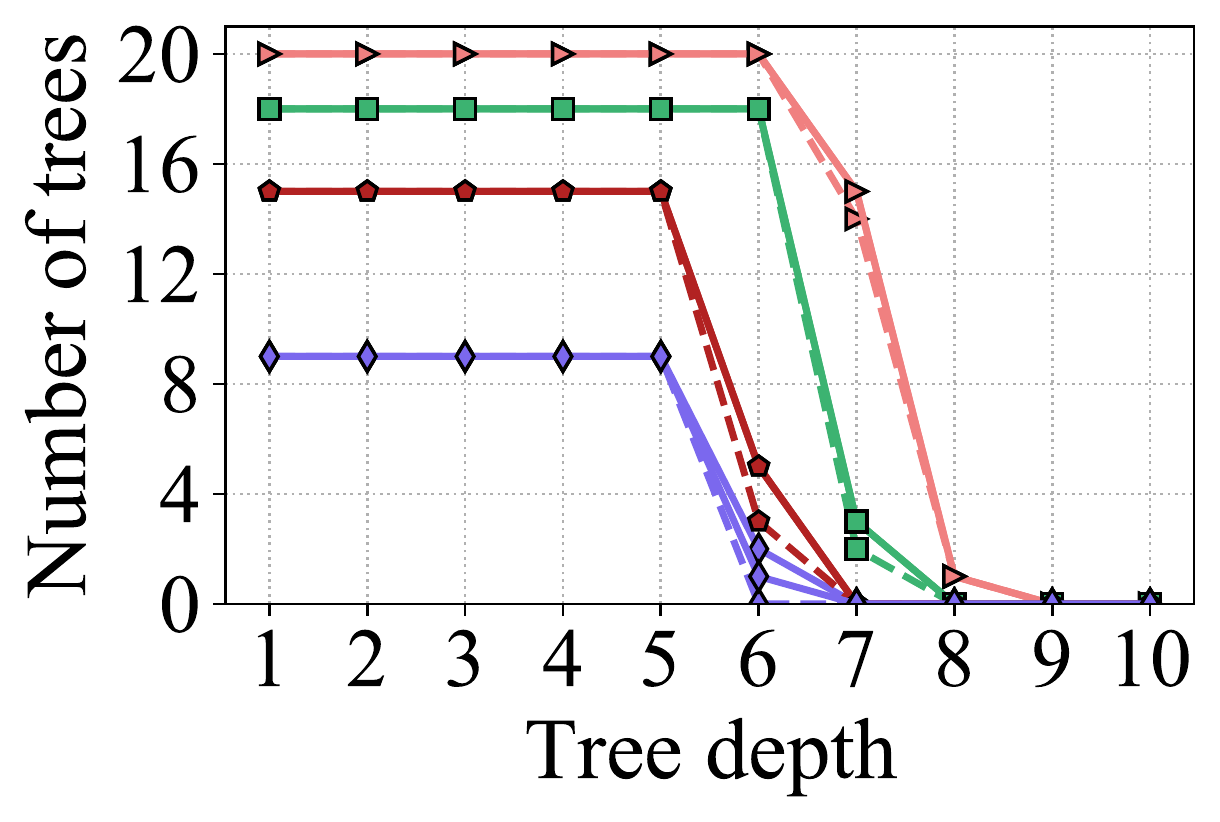}\vspace{-0.5em}\\\centering(d) Financial Transactions.
\end{minipage}
\vspace{-0.5em}
\includegraphics[width=1\linewidth]{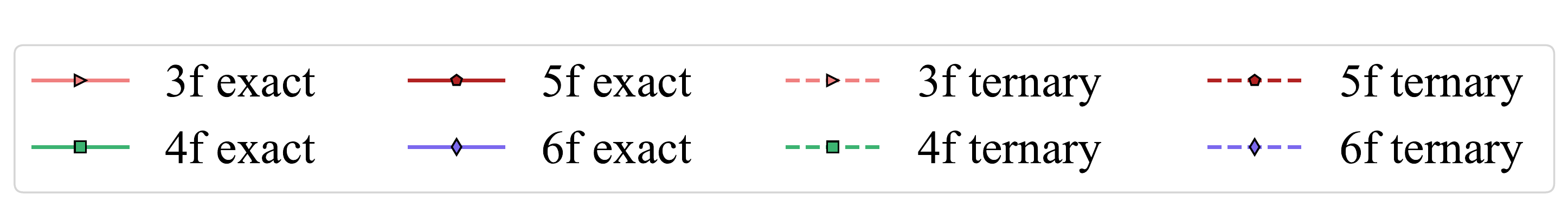}
	\vspace{-1.5em}
	\caption{Ensemble scaling with the number of features.}\label{fig:ensemble_size}
\end{figure}

The size of a model that can be fit within a switch depends not only on the type of the model, but also on the dataset and its features. This is demonstrated in Figure~\ref{fig:ensemble_size} (a) \& (b), which shows how memory requirements of a decision tree scale with the number of features for each of the use cases. These requirements depend on the depth of the feature tables, as well as on the depth of the decision table. In the finance use case, all the features are similar, and adding another feature increases the memory requirements roughly in a consistent manner. In the anomaly detection use case, on the other hand, features vary significantly in their memory requirement. For example, protocol type requires significantly less entries than source or destination port. Consequently, the anomaly detection use case requires less memory than financial transactions. 

\begin{figure}
	\centering
	\begin{minipage}{1\linewidth}
	\includegraphics[width=1\linewidth]{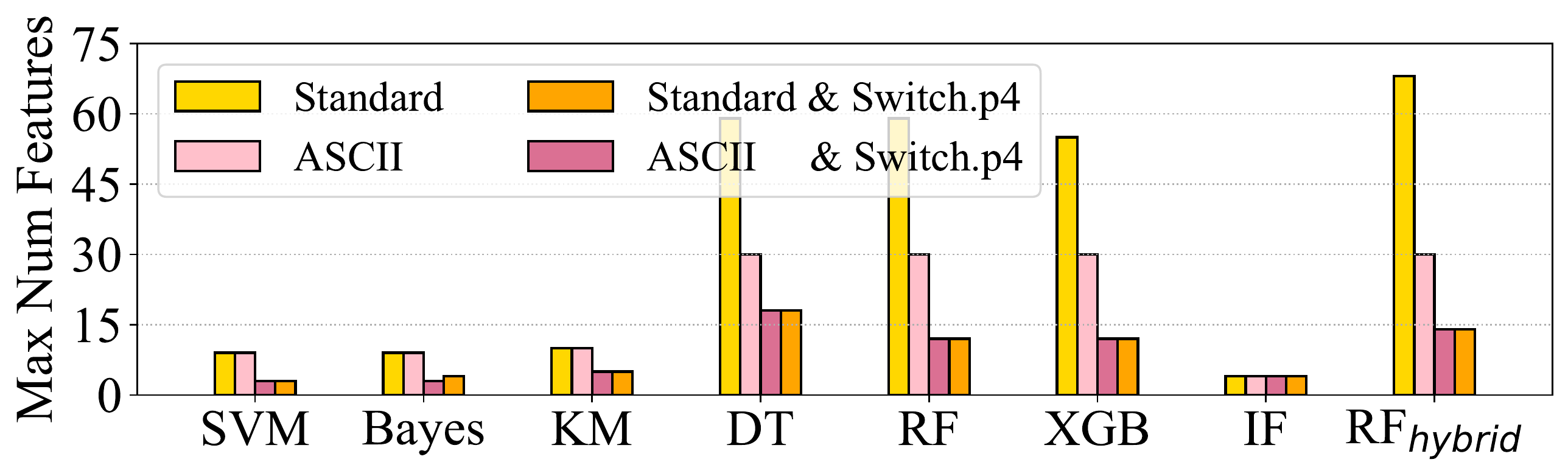}
\end{minipage}
 \vspace{-1em}
	\caption{The maximum number of features available in all IIsy's supported ML models under the financial transaction use case.}\label{fig:max_features}
	\vspace{-1em}
\end{figure}

Fitting an ensemble model within a pipeline requires attending to several constraints: available memory (i.e., table size), number of stages in a pipeline, size constraints on metadata and lookup keys, and logic resources. The overall number of tables used is a soft constraint; as parallel look-ups may happen within the same stage, the overall number of stages is a stronger constraint. 

The difference in the model size that can be fit is demonstrated in Figure~\ref{fig:ensemble_size}. The figure shows the number of trees that can be fit for each of the use cases, as well as their depth, depending on the number of features and the type of memory used (exact match, TCAM or a mix of both). As the figure shows, using up to 6 features, one can fit up to 20 trees. Increasing tree depth means that fewer trees can fit within the switch, due to the size of the decision table. Figure \ref{fig:max_features} shows the maximum number of features allowed in the Tofino pipeline under four implementation variations. Tree models are usually able to utilise more features compare to classical models due to stage sharing. Among these, the standard DT can fit up to 60 features which is due to the number of tables constrained per stage. This number of DT under ASCII implementation is reduced to 30 features, which is because of the depth limitation the parser can parse from the payload. As the results in Table~\ref{tab:ensemble_scalability} show, scaling the number of trees and features has a limited effect on performance, and therefore a smaller switch model may be more efficient when the hybrid model is possible. 
While these results are specific to the switch used, similar constraints exist on different targets.

\begin{table}
	\centering
	\begin{adjustbox}{width=\columnwidth,center}
		\begin{tabular}{|l|c|c|c|c|}
			\multicolumn{5}{c}{Anomaly Detection, Random Forest, 0.7 confidence}\\
			\hline
			&Small & Medium & Large & Baseline\\
			\hline
			Features & 4 & 5 & 6& 25\\
			\hline
			Trees & 6 & 10 & 14& 200\\
			\hline
			Max Depth & 4 & 5 & 6& ---\\
			\hline
			Accuracy & 97.05  & 97.17  & 97.78 & 99.51 \\
			\hline
			Precision & 98.06  & 98.12  &  98.60 &   99.67\\
			\hline
			Recall &  88.55 &  89.04 & 91.36  &   99.75\\
			\hline
			F1 score &  92.60 & 92.94  &  94.58 &  98.88\\
			\hline
			Hybrid Accuracy &   98.58 &  98.94 & 99.31 & ---\\
			\hline
			Hybrid F1 & 96.64  &   97.53 &    98.41& ---\\
			\hline
			\multicolumn{5}{c}{Financial Transactions, XGBoost, 0.7 confidence}\\
			\hline
			Features & 4 & 5& 6& 130\\
			\hline
			Trees & 6 & 10& 14& 200\\
			\hline
			Max Depth & 4 & 5& 6& --\\
			\hline
			Accuracy &  72.48& 72.65 & 73.73& 77.34 \\
			\hline
			Precision & 68.48 &  68.76 &  70.05&  74.43\\
			\hline
			Recall &66.51 & 65.69 &  68.09& 72.76 \\
			\hline
			F1 score &   67.16&  65.51 &  68.78&  73.43  \\
			\hline
			Hybrid Accuracy & 77.31 & 77.30 &  77.26 & ---\\
			\hline
			Hybrid F1 & 73.41  & 73.43 &73.40 & ---\\
			\hline
		\end{tabular}
	\end{adjustbox}
	\vspace{1em}
	\caption{Scalability of ensemble models and resulting \ml performance.}
	\label{tab:ensemble_scalability}
	\vspace{-2em}
\end{table}

\begin{figure}[htb]
	\centering
	\begin{minipage}{0.66\linewidth}
	\includegraphics[width=0.95\linewidth]{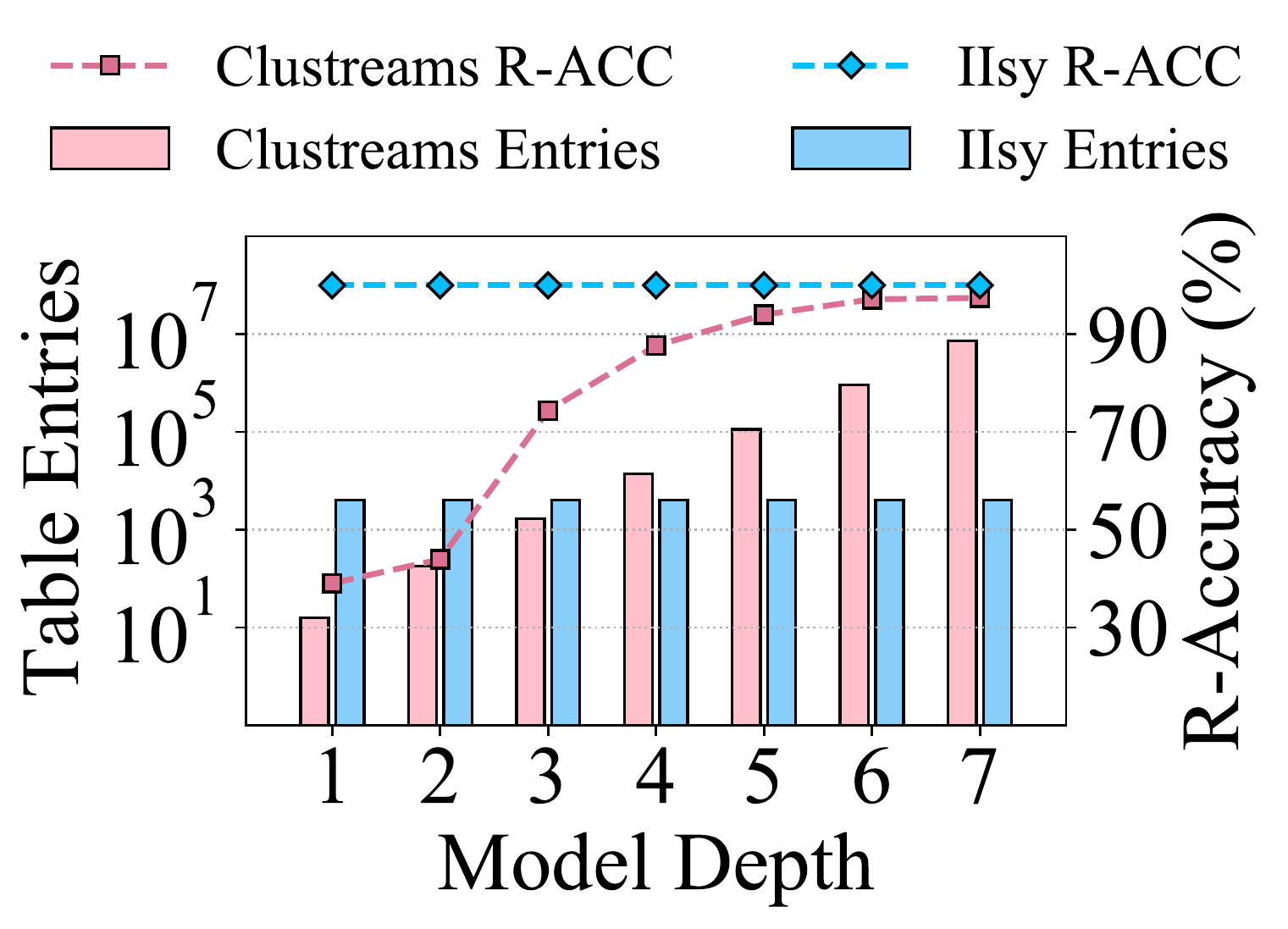} 
\end{minipage}
    \begin{minipage}{0.32\linewidth}
	\includegraphics[width=1\linewidth]{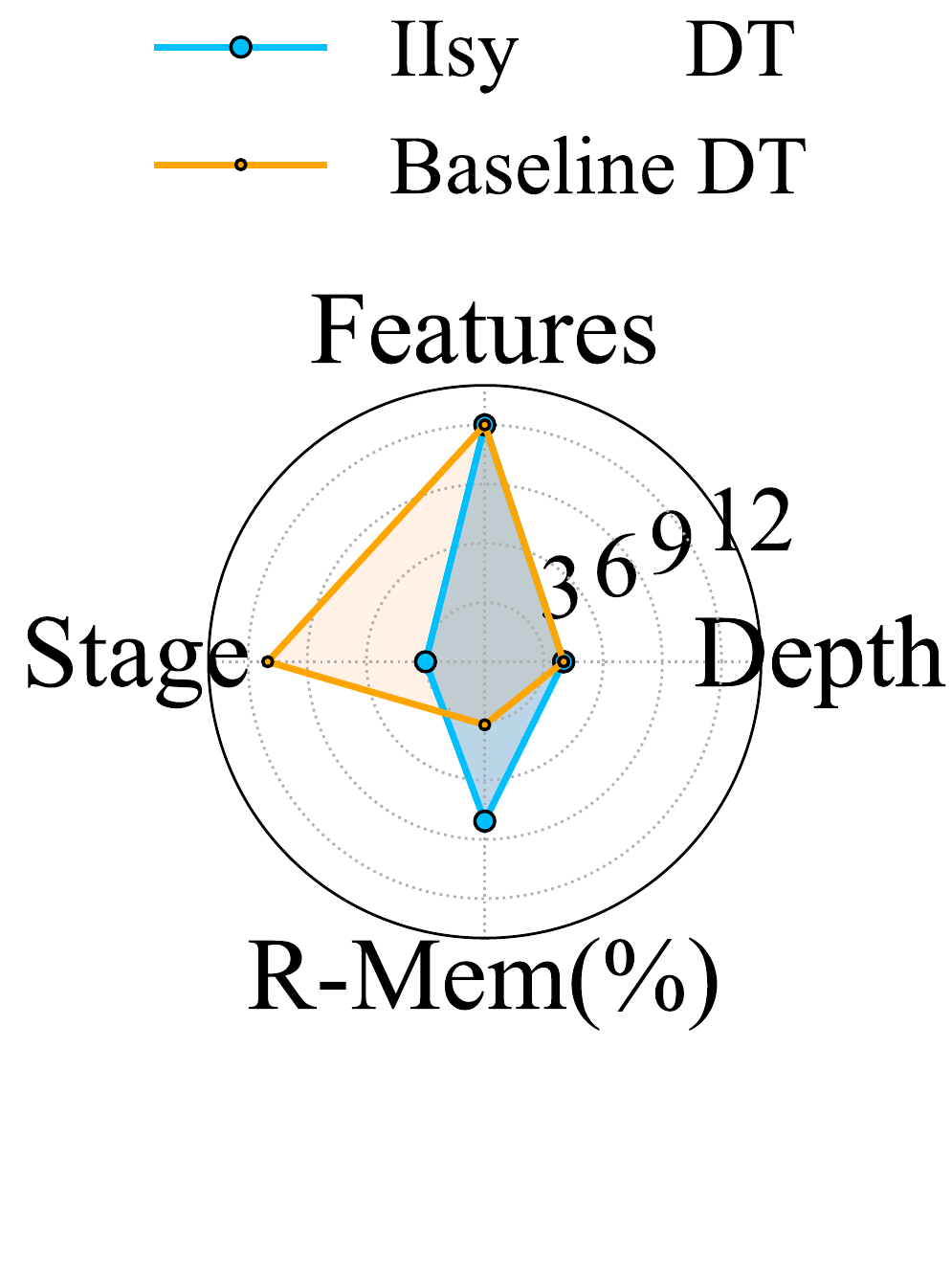}\vspace{-0.5em}  
\end{minipage}\\
\;\;\;\;\;\;\;\;\;\;(a) KM Comparison \;\;\;\;\;\;\;\;\;\;\;\; (b) DT Comparison
 \vspace{-0.5em}
	\caption{Compare to Baseline - (a) IIsy KM compare to Clustreams KM in terms of  Table Entries and (R)elative Accuracy (\%). (b) The IIsy DT compare to SwitchTree DT in terms of Stage and (R)elative (Mem)ory (\%)}\label{fig:km_dt}
\end{figure}

 \begin{figure}[htb]
\includegraphics[width=1\linewidth]{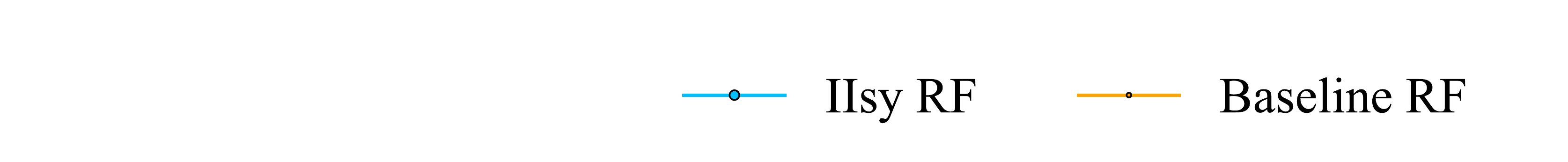}\\
\vspace{0.5em}
\begin{minipage}{0.32\linewidth}
	\includegraphics[width=1\linewidth]{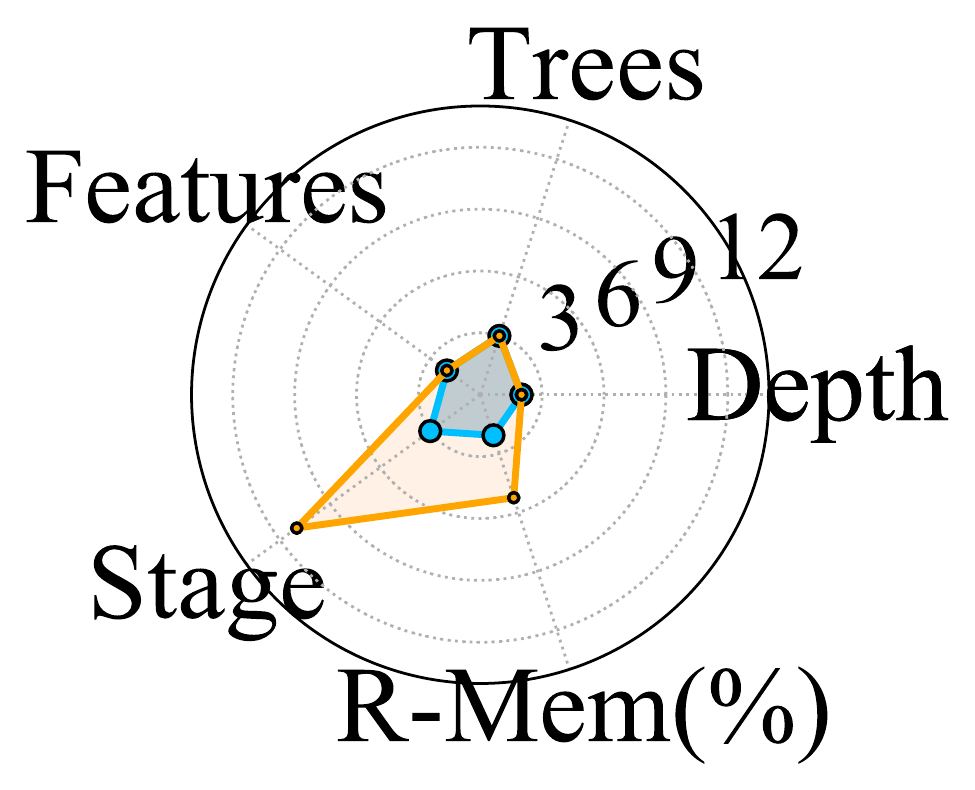} 
\end{minipage}
	\begin{minipage}{0.32\linewidth}
	\includegraphics[width=1\linewidth]{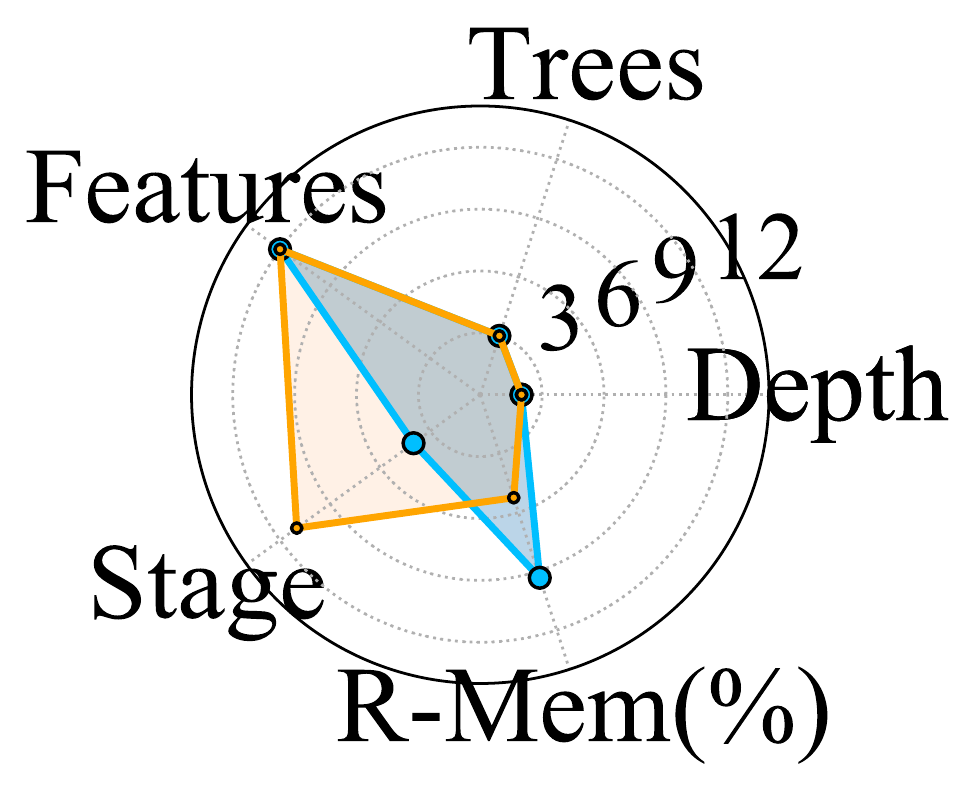} 
\end{minipage}
    \begin{minipage}{0.32\linewidth}
	\includegraphics[width=1\linewidth]{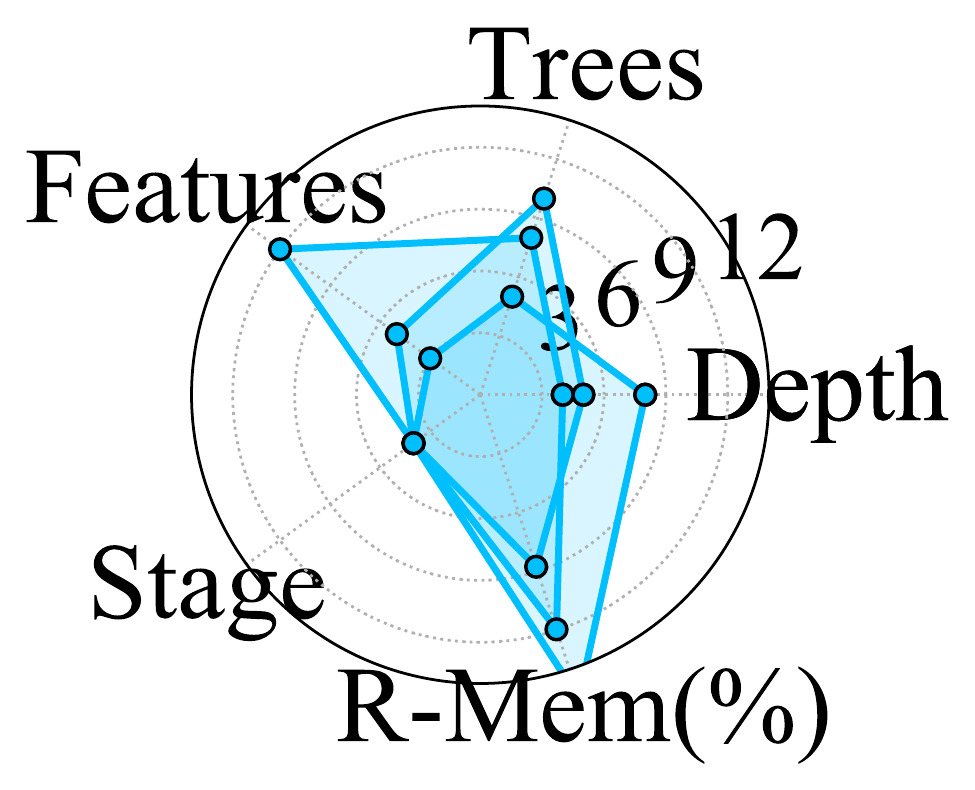} 
\end{minipage}\\
(a) Small Model \;\; (b) Large SwitchTree \; (c) Large IIsy
 \vspace{-0.5em}
	\caption{Compare to Baseline - The (R)elative (Mem)ory and stage consumption of IIsy RF and SwitchTree RF (Baseline) under three sets of hyperparameters.}\label{fig:radar_ensemble_tree}
\end{figure}

\subsection{Baseline Comparison}\label{sec:baseline}

We compare IIsy with two typical in-network Ml research, Clustreams and SwitchTree, in terms of resource consumption under financial transaction use case. As shown in Figure \ref{fig:km_dt}, to achieve the same accuracy, the IIsy's K-means algorithm requires significantly less memory consumption compared to Clustreams. In Figure \ref{fig:km_dt} (b), compared to the SwitchTree, IIsy's DT implementation requires significantly fewer stages (save 8) with only 5\% more relative memory. When it comes to the ensemble model (i.e. RF), as shown in Figure \ref{fig:radar_ensemble_tree} (a), with the small model size, IIsy shows merit in controlling both memory and stage consumption. For the maximum size available for SwitchTree (Figure \ref{fig:radar_ensemble_tree} (b)), in comparison, with only 4\% more memory requirement, IIsy has a significant benefit in stage consumption. In the larger model size (Figure \ref{fig:radar_ensemble_tree} (c)), SwitchTree is unable to map to commodity switches, however, IIsy is able to map and has excellent control over resource consumption.

\subsection{Throughput and Latency}\label{sec:performance}

\begin{figure}[htb]
	\centering
	\begin{minipage}{0.49\linewidth}
	\includegraphics[width=1\linewidth]{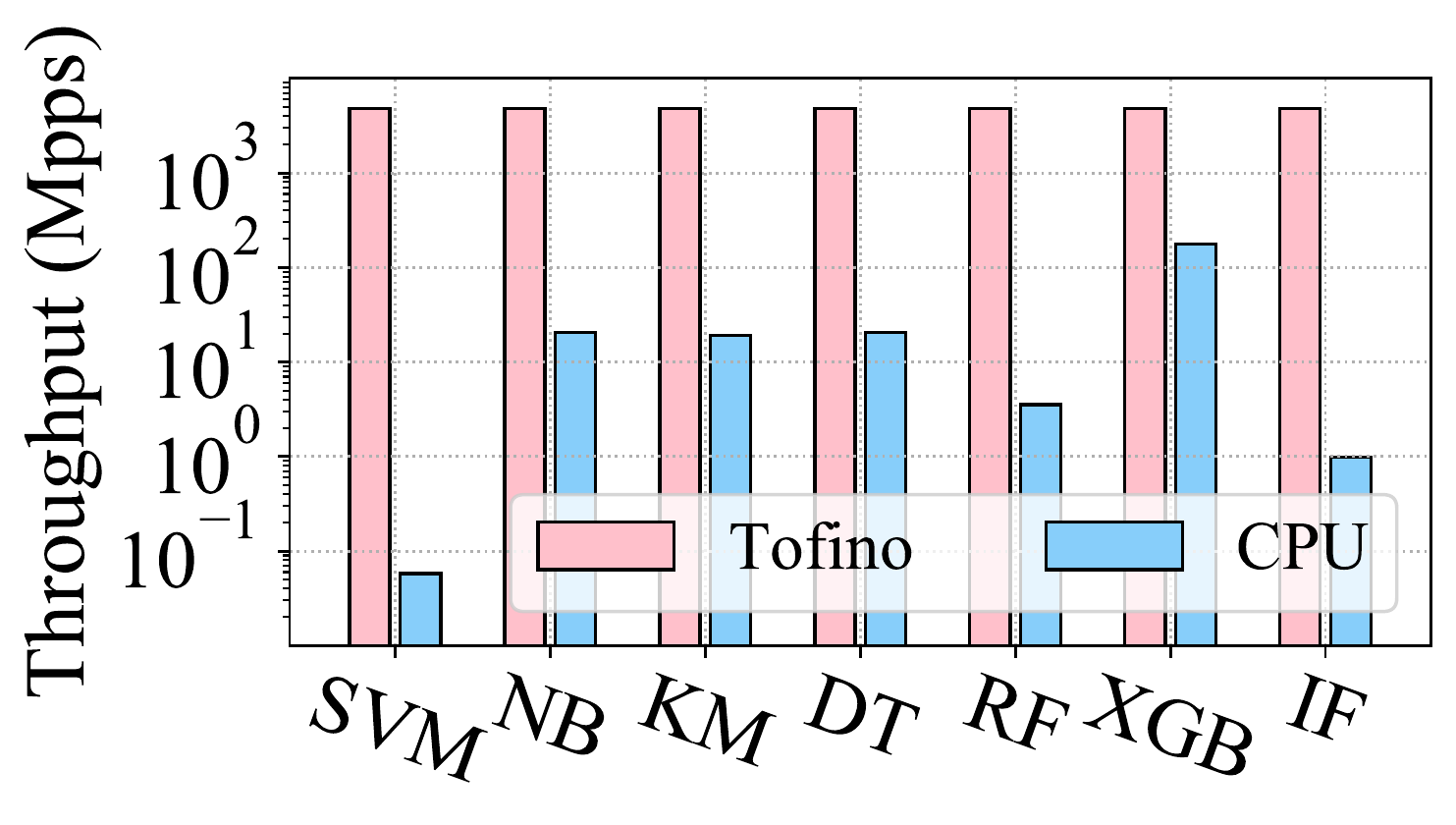}\vspace{-0.5em}\\\centering(a) Throughput
\end{minipage}
	\begin{minipage}{0.49\linewidth}
	\includegraphics[width=1\linewidth]{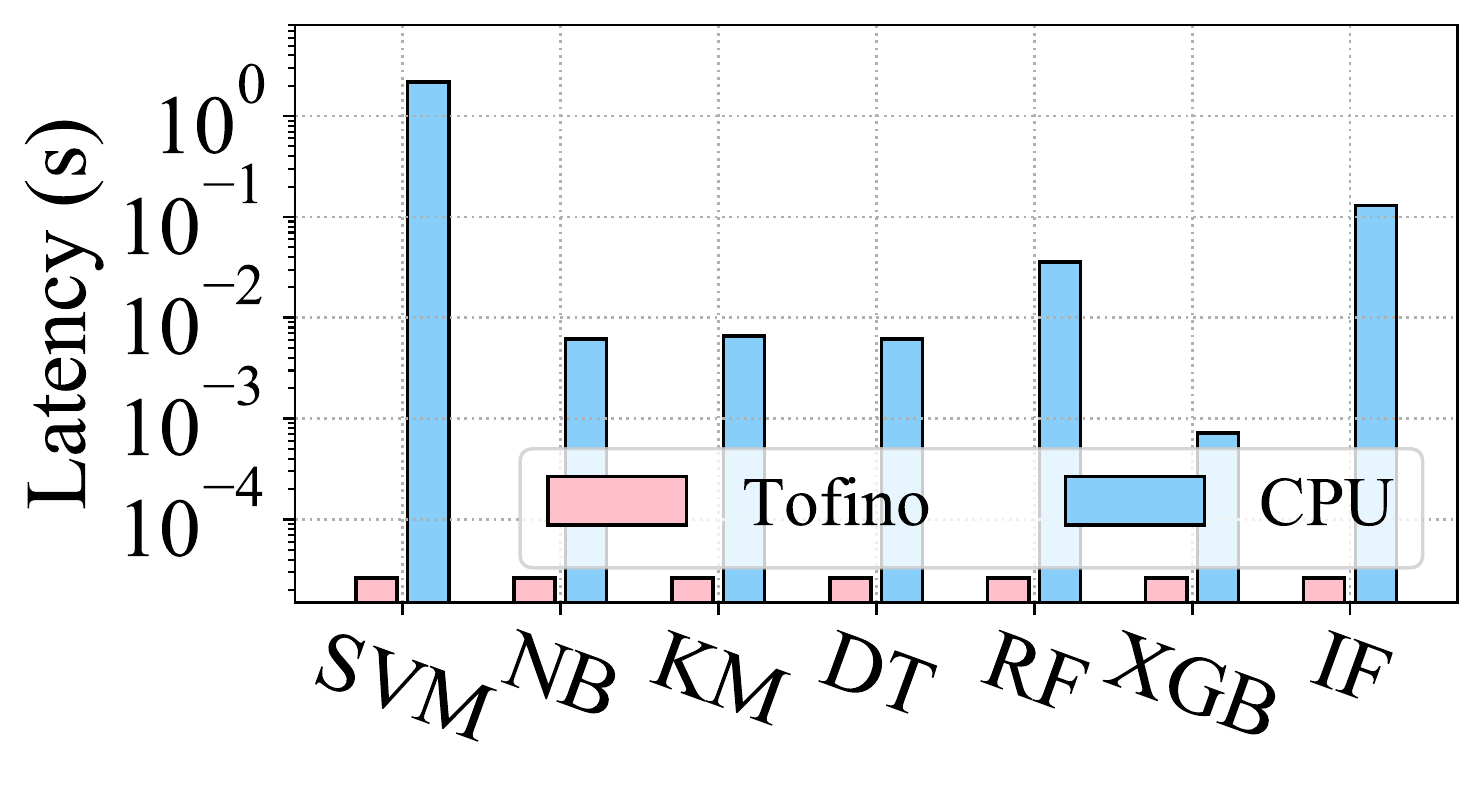}\vspace{-0.5em}\\\centering(b) Latency.
\end{minipage}
\vspace{-1em}
	\caption{Throughput and latency of ML algorithms on Tofino and CPU under financial transaction use case.}\label{fig:throughput_latency}
	\vspace{-1em}
\end{figure}

For both use cases, and for all models, the programs are designed to meet to line rate, with no recirculations or packet drop. The programs meet Tofino's timing for a minimum packet size of 100Gbps per port.
For the anomaly detection scenario, we use UNSW's pcap traces~\cite{moustafa2015unsw}, and for the financial transactions, we used the test dataset and send it over UDP. In both cases, we measure identical throughput to running the same traces through a simple layer 2 forwarding program, with no packet drop on any of the switch's 64 ports. Observe the throughput of the financial transaction use case, Figure \ref{fig:throughput_latency} (a), which shows that the switch implementation achieves a 25 to 80000-fold throughput improvement over the CPU implementation.

The latency of a design is the number of clock cycles in the pipeline reported by the switch compiler, compared those to the switch.p4 reference design. 
For anomaly detection, the latency of the mapped models in the ingress pipeline is between 27.4\% and 45.9\% of the reference design. 
This means that mapping a model to a switch will have a negligible effect on latency if a switch is deployed in its native usage model (\S\ref{sec:model})\footnote{We do not claim to reduce the overall latency of the switch}. A switch acting as an accelerator will add latency to the end-to-end traversal time at the scale of a microsecond~\cite{arista7170white}. Compared with the latency of current financial trading systems~\cite{baron2019risk}, as shown in Figure \ref{fig:throughput_latency} (b), this will save an order of magnitude in latency. Operating on a stock exchange feed, the action on a classified packet can be the actual buy/sell order packet.

\subsection{\ml performance}\label{sec:eval-ml}

We explore the performance of \inl for two scenarios: the classification is done solely in the switch, and the hybrid model. 
In the anomaly detection use case, the dataset is biased, meaning that most of the traffic is normal traffic.  Although SVM, Na\"{\i}ve Bayes, and K-Means achieve a precision of 0.76--0.91 and F1 score of 0.81--0.85, they classify most of the anomaly traffic as normal. To correctly identify anomalies, we focus on the ensemble models. 
In the finance use case,  SVM, Na\"{\i}ve Bayes, and K-Means and achieve a precision of 0.70--0.73 and F1 score of 0.64--0.71.  

\begin{figure}[htb]
	\centering
	\begin{minipage}{1\linewidth}
	\includegraphics[width=1\linewidth]{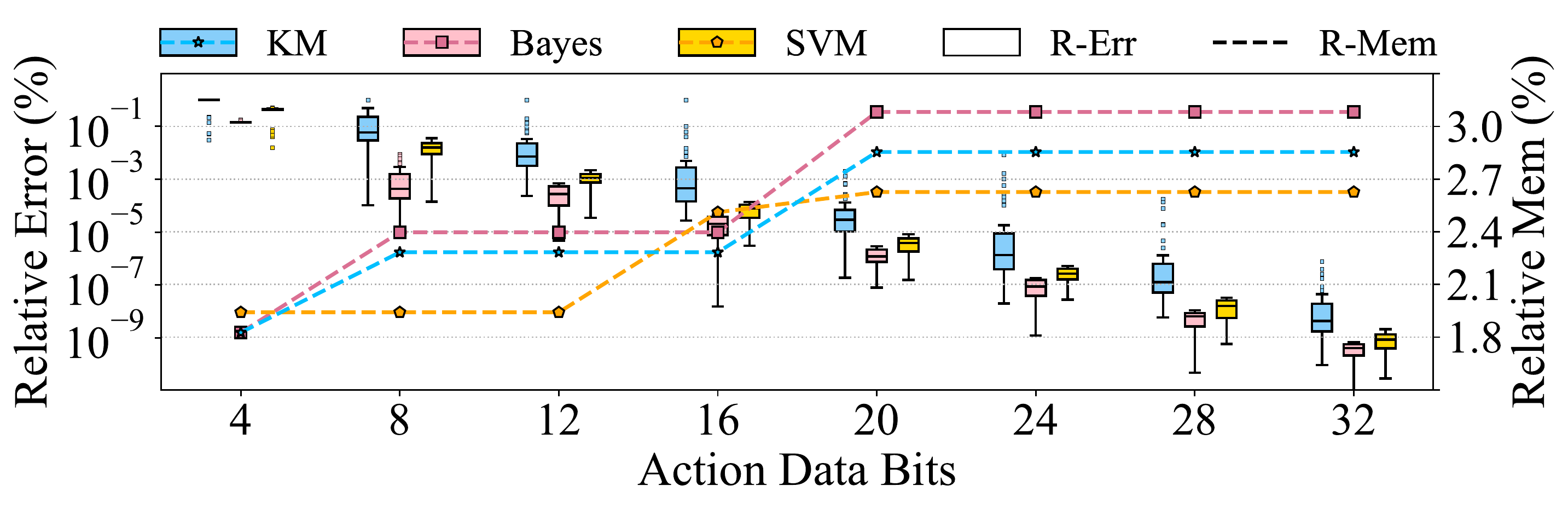}
\end{minipage}
 \vspace{-1em}
	\caption{Calculation error in SVM, Bayes and K-Means.}\label{fig:error}
	\vspace{-1em}
\end{figure}

Our implementations of non-tree based models may introduce an error. We study this error on two front: calculation error and classification error. The calculation error, shown in Figure~\ref{fig:error} for the anomaly detection use case, is the relative error of a result calculated on a switch (e.g., hyperplane equation in SVM), compared with the same equation calculated on a server. While this error is small (less than 0.001\%), the more important result is the misclassification due to calculation error: zero for SVM and K-Means, and 0.00003\% for Na\"{\i}ve Bayes when action data bits is 16. This error is due to extremely low probabilities, and is further eliminated by coding the results of Na\"{\i}ve Bayes calculations, rather than normalizing values. Moreover, as shown in Figure~\ref{fig:error}, the increase in action data bits has a minor effect in terms of memory consumption but significantly reduces the calculation error. 

Next, we consider the use of a hybrid deployment. The baseline is the full ensemble model running on back-end servers. We implement on the switch a smaller model that classifies a subset of the traffic, where all traffic not classified or classified with low confidence goes to the back-end as before. 
A confidence level is set in the switch to determine the threshold for classification on the switch.
The performance of the small model, running on a switch alone, compared with the full model, is shown in Table~\ref{tab:ensemble_scalability}. 

\begin{figure}[t]
	\centering
	\begin{minipage}{0.49\linewidth}
	\includegraphics[width=1\linewidth]{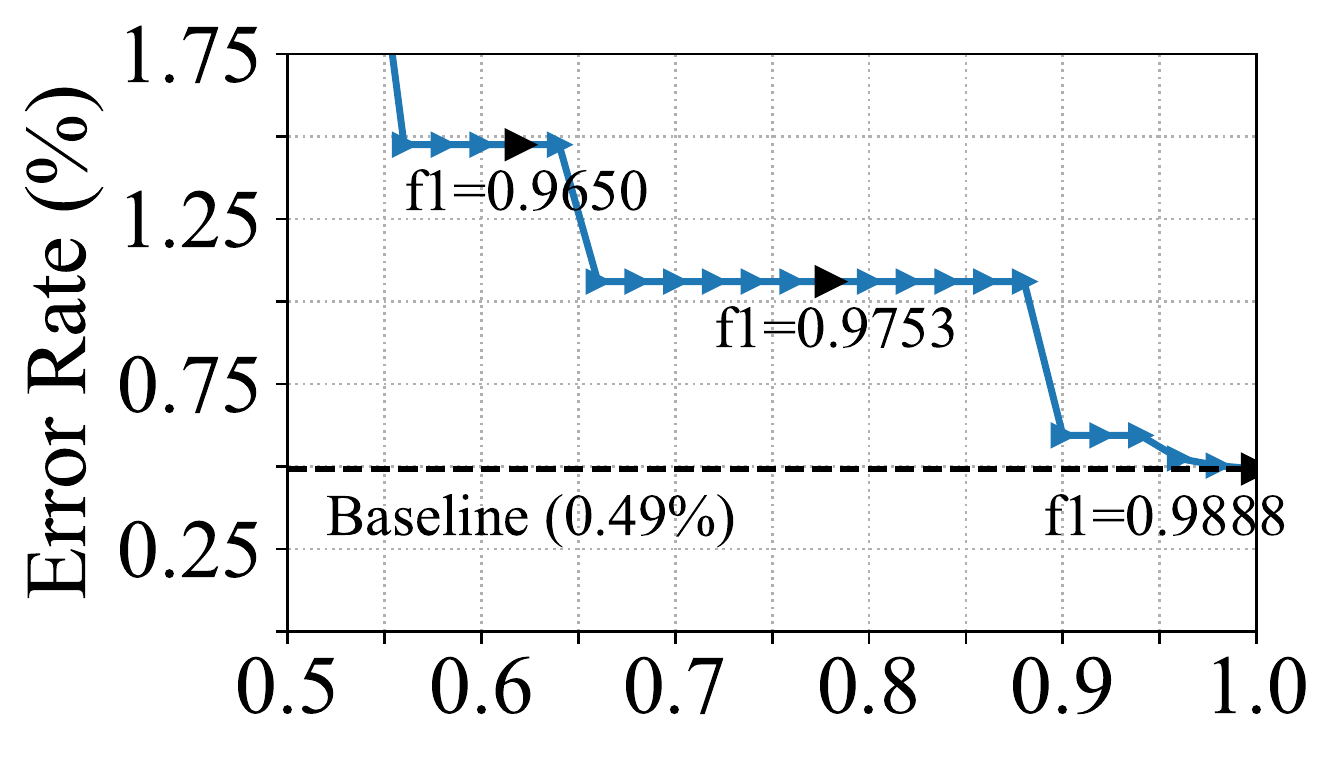}\vspace{-0.5em} 
\end{minipage}
	\begin{minipage}{0.49\linewidth}
	\includegraphics[width=1\linewidth]{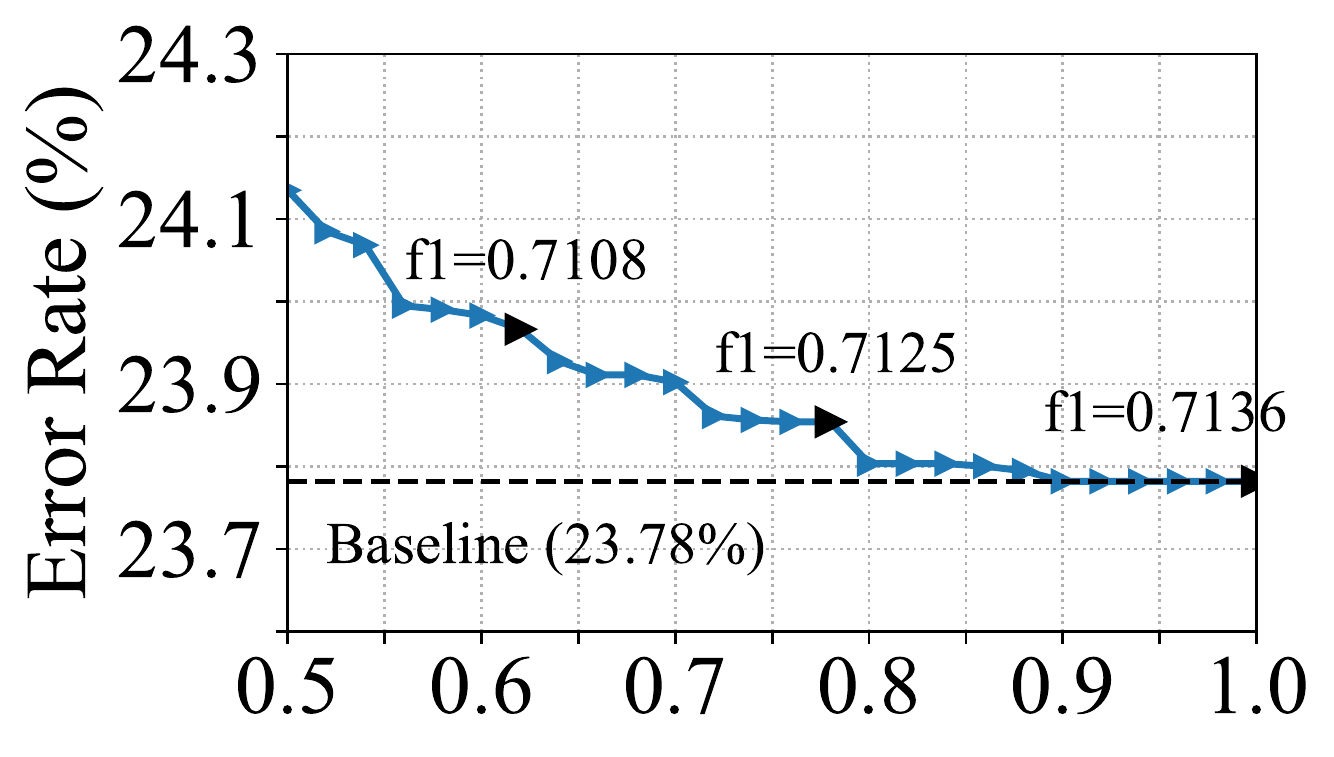}\vspace{-0.5em} 
\end{minipage}
\vspace{0.1em}

\begin{minipage}{0.49\linewidth}
	\includegraphics[width=1\linewidth]{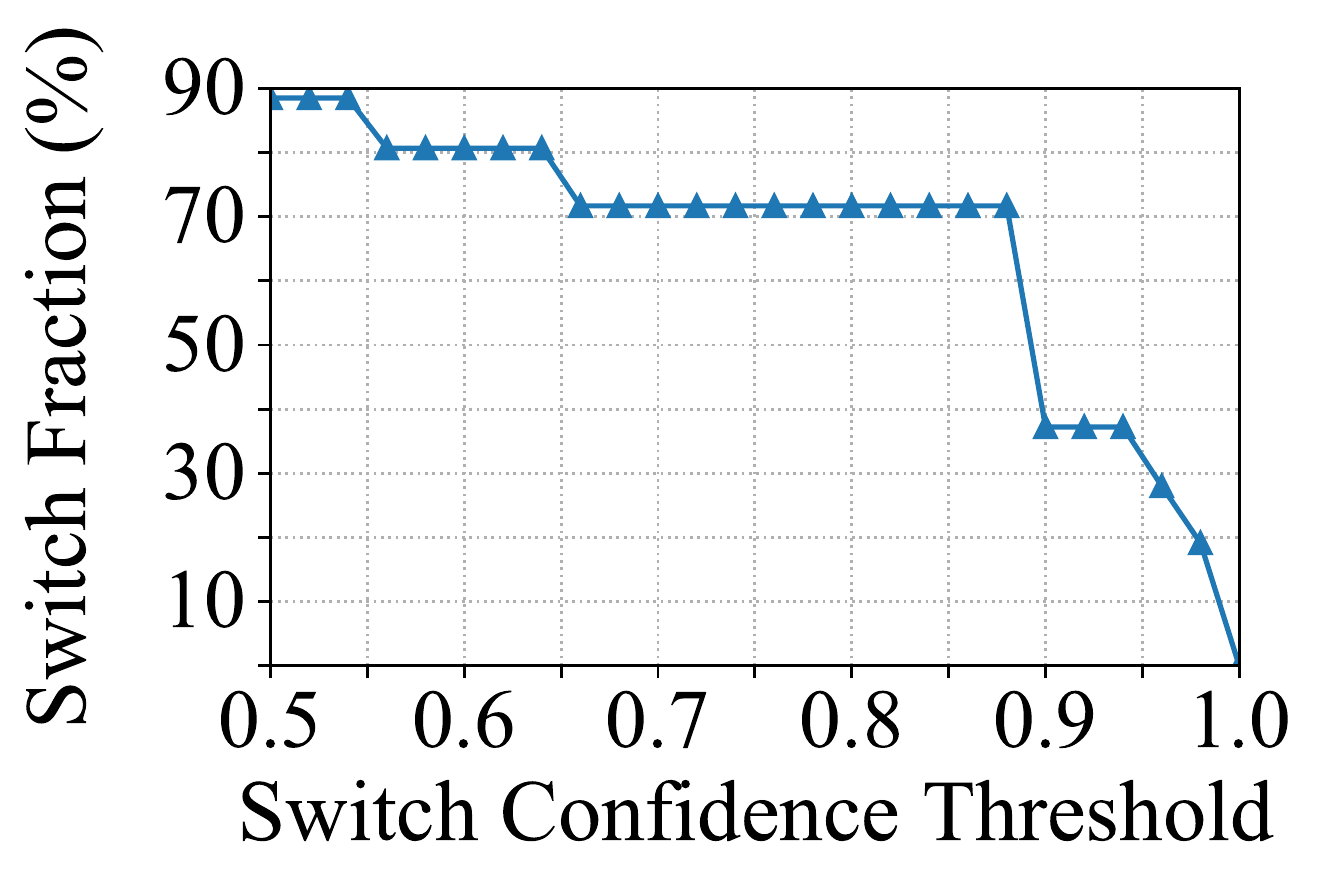}\vspace{-0.5em}\\\centering(a) Anomaly detection
\end{minipage}
	\begin{minipage}{0.49\linewidth}
	\includegraphics[width=1\linewidth]{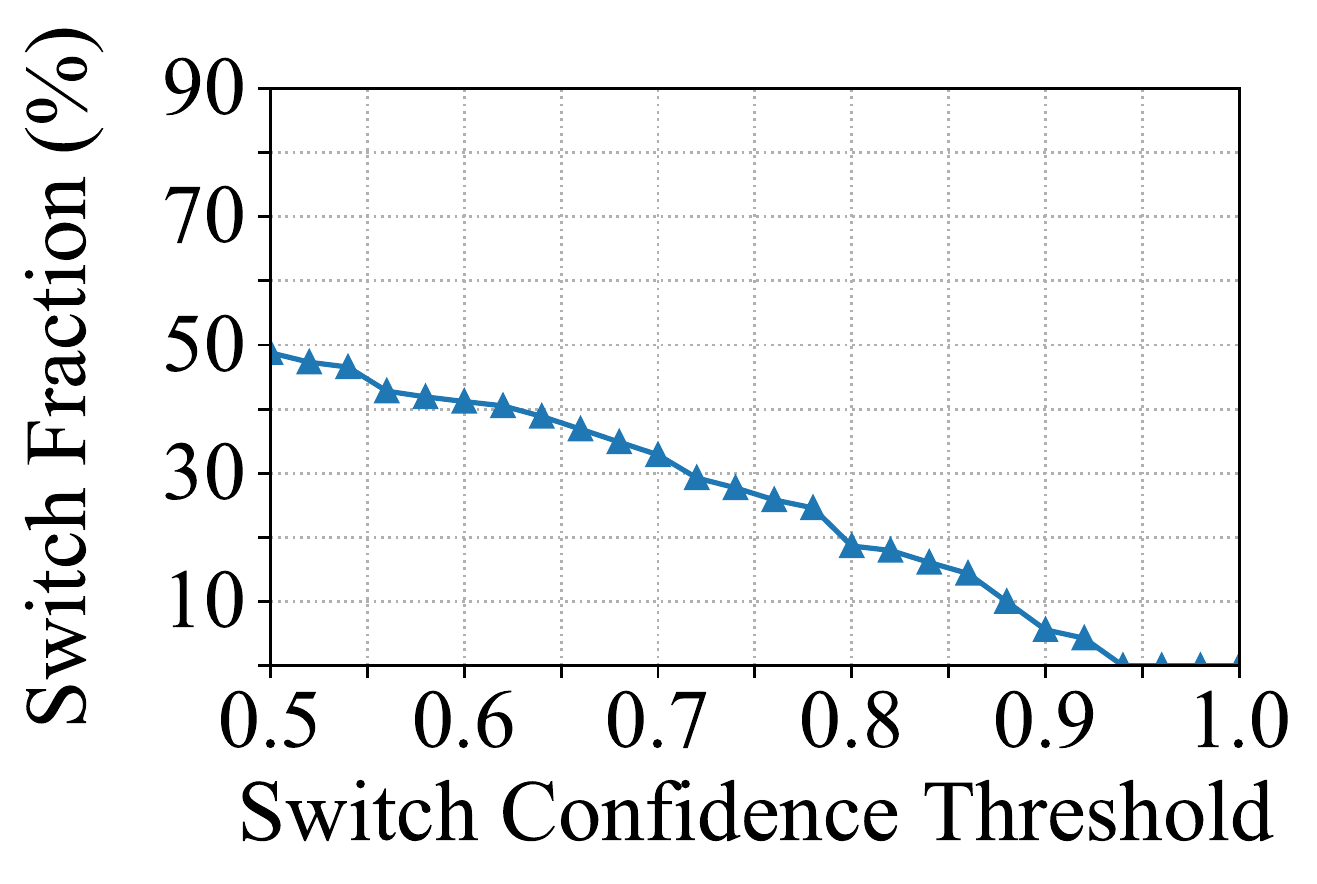}\vspace{-0.5em}\\\centering(b) Financial transactions.
\end{minipage}
	\vspace{-0.5em}
	\caption{Fraction of traffic handled by the switch and misclassification rate.}\label{fig:unsw_ml}
\end{figure}

Figure~\ref{fig:unsw_ml} (a) shows for the anomaly detection use case using Random Forest, the fraction of traffic offloaded by the switch and the corresponding misclassification rate, as a function of the switch classification confidence threshold.  
The baseline results in a misclassification rate of 0.49\%, and F1 score of 0.9888. In comparison, with a confidence threshold of 0.7, 84.5\% of the traffic is handled by the switch, achieving a misclassification rate of 1.03\% and F1 score of 0.976. These improve as the confidence threshold increases, but the fraction of traffic handled by the switch decreases. Where Figure~\ref{fig:unsw_ml} (b) shows a similar result in Financial transaction use case.


\begin{figure*}[t]
	\centering
    \hspace{-0.9em}
	\begin{minipage}{0.33\linewidth}
	\includegraphics[width=1.1\linewidth]{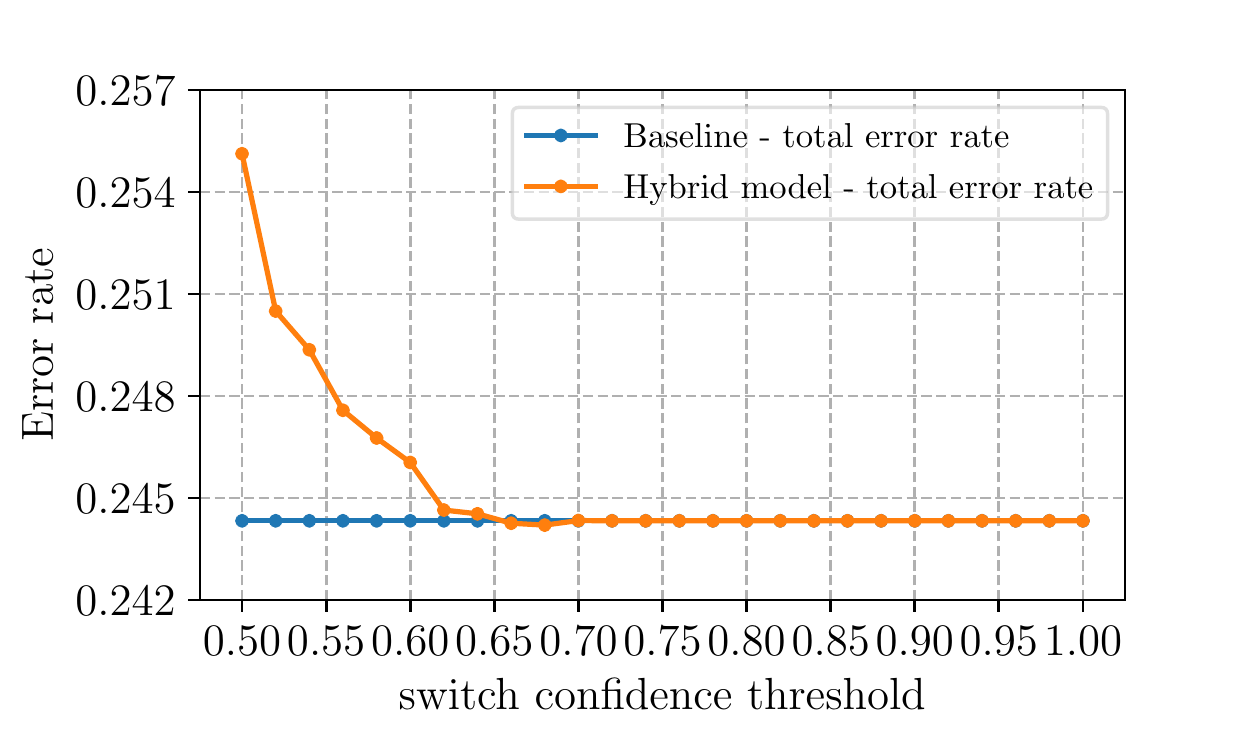}\vspace{-0.5em}\\\centering(a) Confidence to Total Error
	\end{minipage}
	\begin{minipage}{0.33\linewidth}
	\includegraphics[width=1.1\linewidth]{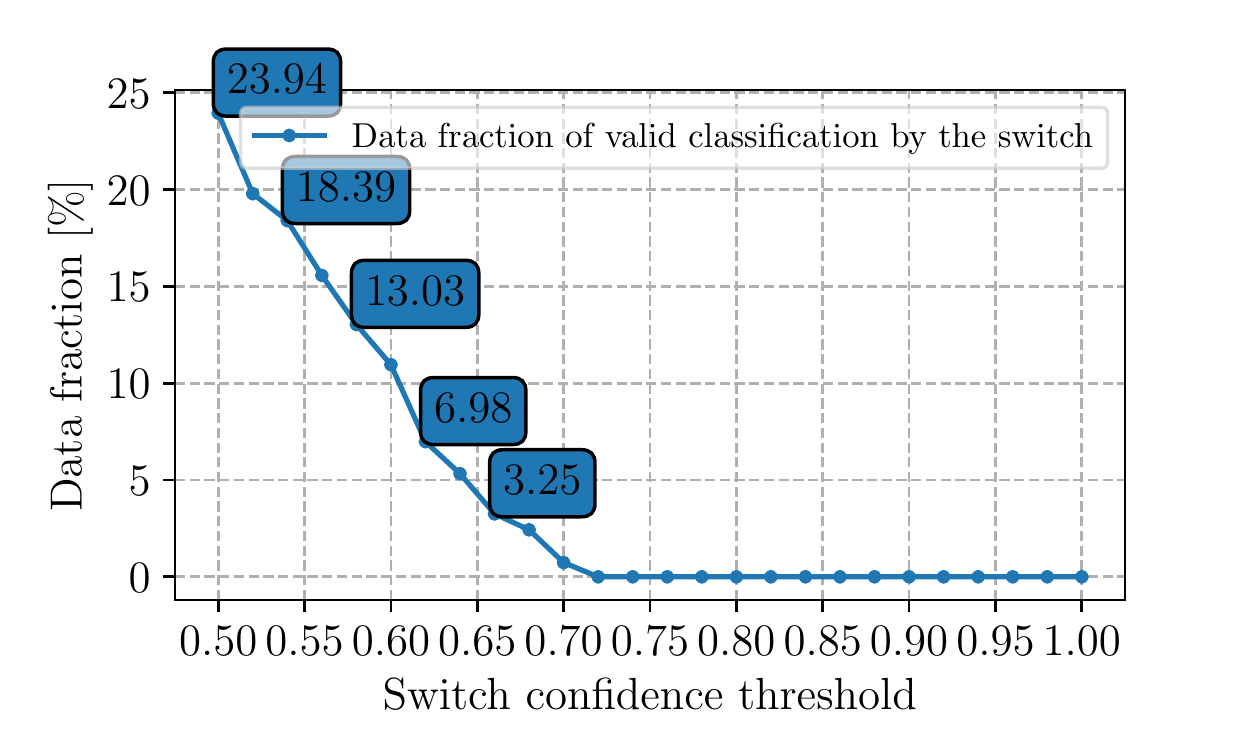}\vspace{-0.5em}\\\centering(b) Confidence to Fraction of Traffic 
	\end{minipage}
	\begin{minipage}{0.33\linewidth}
	\includegraphics[width=1.1\linewidth]{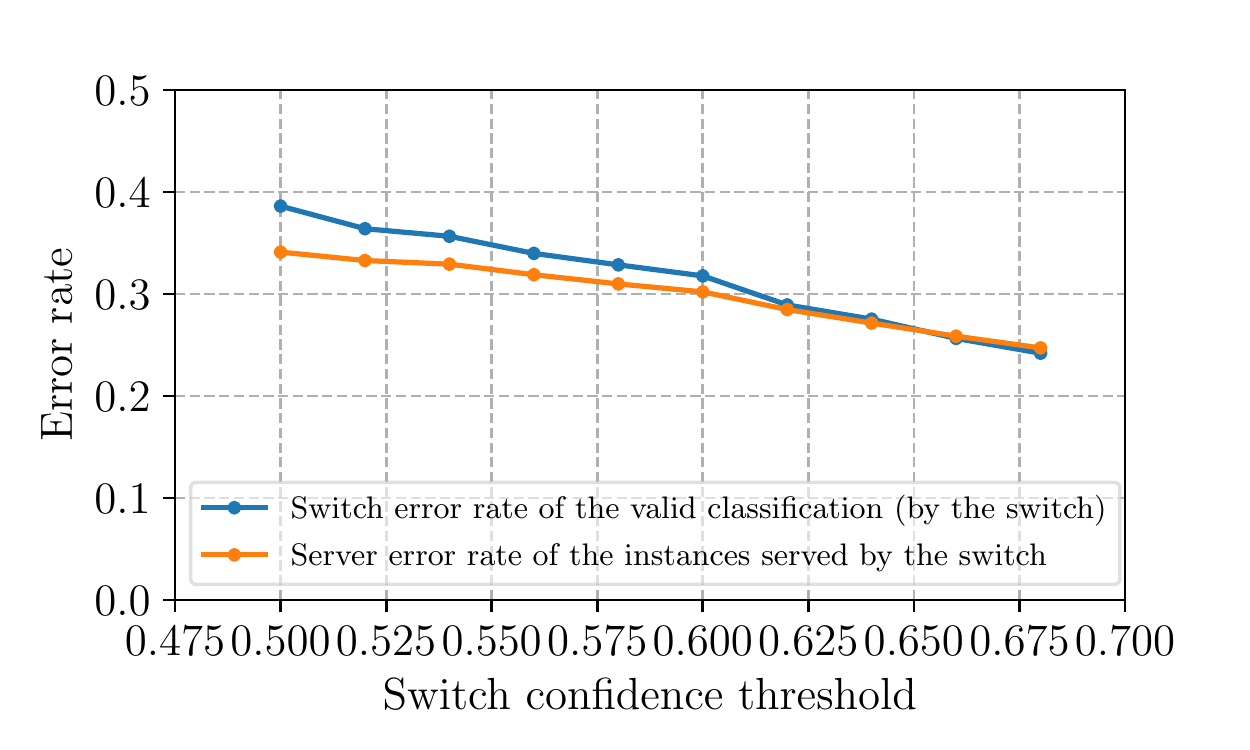}\vspace{-0.5em}\\\centering(c) Confidence to Error Rate
	\end{minipage}
	\caption{Latency sensitive financial transactions: The effect of confidence threshold} 
	\label{fig:finance_ml}
	\vspace{-1em}
\end{figure*}

Figure~\ref{fig:finance_ml} presents the effect of confidence threshold on the performance of financial transactions use case. Figure~\ref{fig:finance_ml})(a) shows that the XGBoost model running in the back-end, using 100 trees with a maximum depth of 8, achieves an error rate of 0.244.
In comparison, the hybrid model achieves an error rate 0.255 with a confidence threshold of 0.5. Increasing the confidence level to 0.65 reduces the error rate to 0.23. However, there is a trade-off here, shown in Figure~\ref{fig:finance_ml})(b): with a threshold of 0.5, 23.9\% of strong buy/sell transactions are classified by the switch, whereas at 0.66 confidence, only 3.25\% of the transactions are classified by the switch. To put these results in context, consider Figure~\ref{fig:finance_ml})(c), which shows the error rate for classifications done by the switch compared with the error rate for the same transactions, if done by the host. As the graph shows, transactions that achieve low confidence (below 0.55) on the switch, are also more likely (over 30\%) to be misclassified by the full-grown model running on the server. In fact, starting 0.6 confidence threshold (where 10.8\% of strong sell/ buy are being served by the switch model) the difference in error rate between the server and the switch is very small, and some traders may even find that the error difference at 0.54 is still small enough to provide higher transactions rate for 18.4\% of the transactions.


\subsection{Optimizations}\label{sec:optimize}

The results presented in this section are the output of IIsy. These results can be further optimized by the user according to needs, either by changing IIsy's configurations (in most cases) or manually (rarely). 

The easiest resource optimization for ensemble models is reducing the number of trees or their depth in the training stage, thereby reducing \ml performance. As demonstrated in Table~\ref{tab:ensemble_scalability}, in a hybrid deployment, there is a minor added benefit to using \textit{Medium} or \textit{Large} switch models over a \textit{Small} model. This can further be tweaked by changing the confidence threshold, and is configurable.

Memory resources can be saved by binning, i.e. mapping more entries in the features tables to the same code. This in turn reduces the size of classification tables and tree tables. This may lead to loss of \ml performance and is controlled by a configuration. 

Overall resources, including both memory and the number of tables, can also be reduced by combining feature tables, meaning using as a key to a table the concatenation of multiple features. This currently requires a manual change of one line in the p4 code and a small change in the mapping tool.

\subsection{ML Model Updates}\label{sec:updates}

We measure the update time of our \ml ensemble models on a switch. Our assumption is that due to changes in data over time, the result of the training has changed, but not the type of the model (e.g., XGBoost) or the constraints used for the training (e.g., maximum tree depth). The update time varies based on the size of the model: from $50ms$ for a small model, for several seconds for a large one (Table~\ref{tab:ensemble_scalability}).

\section{Discussion}\label{sec:discussion}

\textbf{Generalization} The focus of this paper is on the methodology of mapping \ml models to network devices. \name uses a simple data plane, with complexity mainly in the algorithms mapping from the trained models to table entries. Porting between targets is straight forward, as was the case in porting between NetFPGA and Tofino. It requires syntax changes in the P4 code generator and a script generating control-plane commands, but not to the mapping tool. Planter~\cite{planterarxiv, zheng2021planter} builds upon IIsy to support more models and targets. 

\textbf{Benefits} A lesson of this work is that despite resource constraints, network switches can  serve as important classification components in hybrid deployments. Saving microseconds (or more) of latency in time-sensitive applications or reducing back-end servers load by tens of percent, without adding new hardware to the infrastructure is a key element. While classification can not be implemented within a fully utilized switch, our results show that the resource overheads of adding classification functionality to a switch are practical. 

\textbf{Scope} This paper focuses on mapping trained \ml models to network devices. The work does not seek to improve the quality of training \ml models, nor to contribute to a specific use case. 
Applying the methodology to certain applications, such as congestion control, is beyond the scope of the paper. 
While the methodology offered in this work is can not be directly applied to neural networks models, our choice of ensemble models is because they provide the best results for the  example use cases.

\textbf{Egress Pipeline} Using the egress pipeline is unlikely in most use cases, as typically the output port is determined in the ingress pipeline.  While we explored using an egress pipeline, it did not improve the scalability of IIsy.

\textbf{Limitations} Some of the limitations discussed in this work, e.g., the number of tables or features, are property of the target platform and will change on a different platform. For example, NetFPGA is mostly limited by memory and logic resources, while on Tofino, memory and logic resources are rarely limiting us, and we are limited by different constraints, such as the number of stages.

\section{Related work}\label{sec:related}

\begin{table}
	\begin{adjustbox}{width=\columnwidth,center}
		\centering
		\begin{tabular}{lccccc}
			\toprule
			Project  & Target & Models & OTS & Const. & OA\\
			\hline
			\grayrow BaNaNa~\cite{sanvito2018can,siracusano2018network} & RMT, NIC & BNN & \cmark & P &  \xmark\\
			N3IC~\cite{siracusano2020running,siracusano2022re} &  NIC, FPGA & BNN & \cmark & P & \xmark\\
			\grayrow Qin~\cite{qin2020line} & bmv2, NIC & BNN & \cmark & \xmark  & \cmark\\
			\hline
			IOI\cite{zhong2021ioi} & ASIC & NN & \xmark & --- & \xmark  \\
			\grayrow iSwitch~\cite{li2019accelerating}  & FPGA & RL & --- & P &  \xmark\\
			Taurus~\cite{swamy2022taurus} & ASIC & DNN, SVM,  & \xmark & \cmark &  \xmark\\
			&  &  KM, LTSM &  &  &\\
			\hline
			\grayrow pForest~\cite{busse2019pforest} & bmv2, ASIC & RF & \cmark & \cmark & \xmark\\
			SwitchTree~\cite{lee2020switchtree} & bmv2 & RF & --- & \xmark &  \cmark\\
			\grayrow         NERDS\cite{xavier2021programmable} & bmv2, NIC & RF & \cmark& P &  ?\tablefootnote{The repository linked by the authors is not available at this time}\\
			\hline
			 Clustreams~\cite{friedman2021clustreams} & ASIC & KM & \cmark & \xmark  & \xmark\\
			\hline
			IIsy & ASIC,  & SVM, KM, NB  & \cmark & \cmark & \cmark\\
			&  FPGA&   RF, XGB, IF &  &    &\\
			\bottomrule         
		\end{tabular}
	\end{adjustbox}
	
	\vspace{1em}
	\caption{ A comparison of \inl solutions. Legend: OTS - Off the shelf. Const. - Resource constrained. OA - Open Access. NN - Neural Network. BNN/DNN - Binary/Deep NN. RF - Random Forest. NB -  Naïve Bayes. KM- K-Means. XGB - XGBoost. IF - Isolation Forest. P - Partial.}
	\label{tab:related_comparison}
	\vspace{-2em}
\end{table}

The application of \ml to network traffic, and in particular the use of \ml for traffic classification, has been of an interest for a long time (e.g., ~\cite{moore2005internet,dainotti2012issues}). Using \ml for scheduling and congestion control (e.g.,~\cite{dhukic2019advance}) was also studied. The focus of most of these works has been on using \ml over traditional computing platforms.


The challenges of \ml have led researchers to explore new approaches to resource constrained \ml, using devices of limited resources~\cite{wang2019adaptive}. Such approaches are popular with IoT devices (e.g.,~\cite{sanchez2020tinyml}). This work focuses on network switches, which have more resources than some of these devices, and also much higher processing rate and a different architecture.

A related thread of research is using programmable switches to accelerate \ml frameworks. These works focus on parameters servers and in-network aggregation~\cite{lao21atp,sapio21switchml} in the training stage, rather than the classification.

A few works tried to implement \ml models within network devices. Implementing binary neural networks was explored in N2Net~\cite{siracusano2018network} and BaNaNa Split~\cite{sanvito2018can}, with limited performance benefits. pForest~\cite{busse2019pforest}, SwitchTree~\cite{lee2020switchtree} and NERDS~\cite{xavier2021programmable} explored mapping random forests, with only pForest attempting ASIC implementation. Their methodology is different, encoding each decision tree in separate tables, with a table for every tree level, achieving lower scalability (e.g., depth of 4 in pForest). Li~\cite{li2019accelerating} added an acceleration module within a switch for reinforced learning, and Taurus\cite{swamy2022taurus} suggested adding a map-reduce module. Taurus did not report \ml performance results for the models supported by \name. IIsy's use of unmodified network devices is complementary to these works.

Using programmable network devices for anomaly detection was explored both at the host side~\cite{zhao2020achieving} and within switch-ASIC (e.g.,~\cite{liu2021jaqen}). Our work is orthogonal to these non-\ml based efforts, as it enables \ml based anomaly detection solutions (e.g., \cite{injadat2018bayesian}) to be migrated to the network. 

\ml for financial transactions has been widely researched, with XGBoost and SVM often used~\cite{tsantekidis2017using}. Acceleration of financial transactions has mostly focused on the back-end, e.g. using FPGA~\cite{velu2020algorithmic}. The closest programmable switches project is the publish-subscribe system~\cite{jepsen2020forwarding}, for the NASDAQ Market data feed filter and router.

\section{Conclusion}\label{sec:conclusion}

The toll of running \ml workloads is high. In this paper, we have made the case for \inl. By mapping multiple \ml models, including ensemble models, to off-the-shelf network switch, we have demonstrated the feasibility and benefits of running classification within network devices. 

This paper complies with all applicable ethical standards of the authors' home institution.\\

\noindent\textbf{Acknowledgements}\label{sec:acknowledgement}
This work was partly funded by VM-Ware, the Leverhulme Trust (ECF-2016-289)
and the Isaac Newton Trust.

\bibliographystyle{ACM-Reference-Format}
\bibliography{ref.bib}

\newpage

\appendix

\section{Mapping Models - Additional Information}\label{app:models}

\subsection{SVM}\label{sec:SVM}

Support vector machines (SVM) use hyperplanes to separate between classes, where the output of the training stage is the equations of the hyperplanes , such as: 

$$\left\{\begin{matrix}
	a_{1}x_{1}+b_{1}x_{2}+...z_{1}x_{n} + d_{1} = 0\\ 
	a_{2}x_{1}+b_{2}x_{2}+...z_{2}x_{n} + d_{2} = 0\\ 
	...\\ a_{m}x_{1}+b_{k}x_{2}+...z_{m}x_{n} + d_{m} = 0\end{matrix}\right.$$

where $x_{i}$ is the value of feature {i}, $n$ is the number of features, $k$ is the number of classes and $m = k*(k-1)/2$. 

There are two ways to map SVM to a network device. First, to hold a table per feature, and second, to hold a table per hyperplane. A table per feature means that the key to a table is the feature's value, and the output of the lookup is a vector of calculated values $a_{i}\times x_{i}$. The value of each hyperplane is calculated as the sum of vectors from all feature tables, and a decision is taken. This can be optimized by adding up the features in each pipeline stage.
\begin{figure}[htbp]
	\centering
	\vspace{2em}
	\includegraphics[width=1\columnwidth]{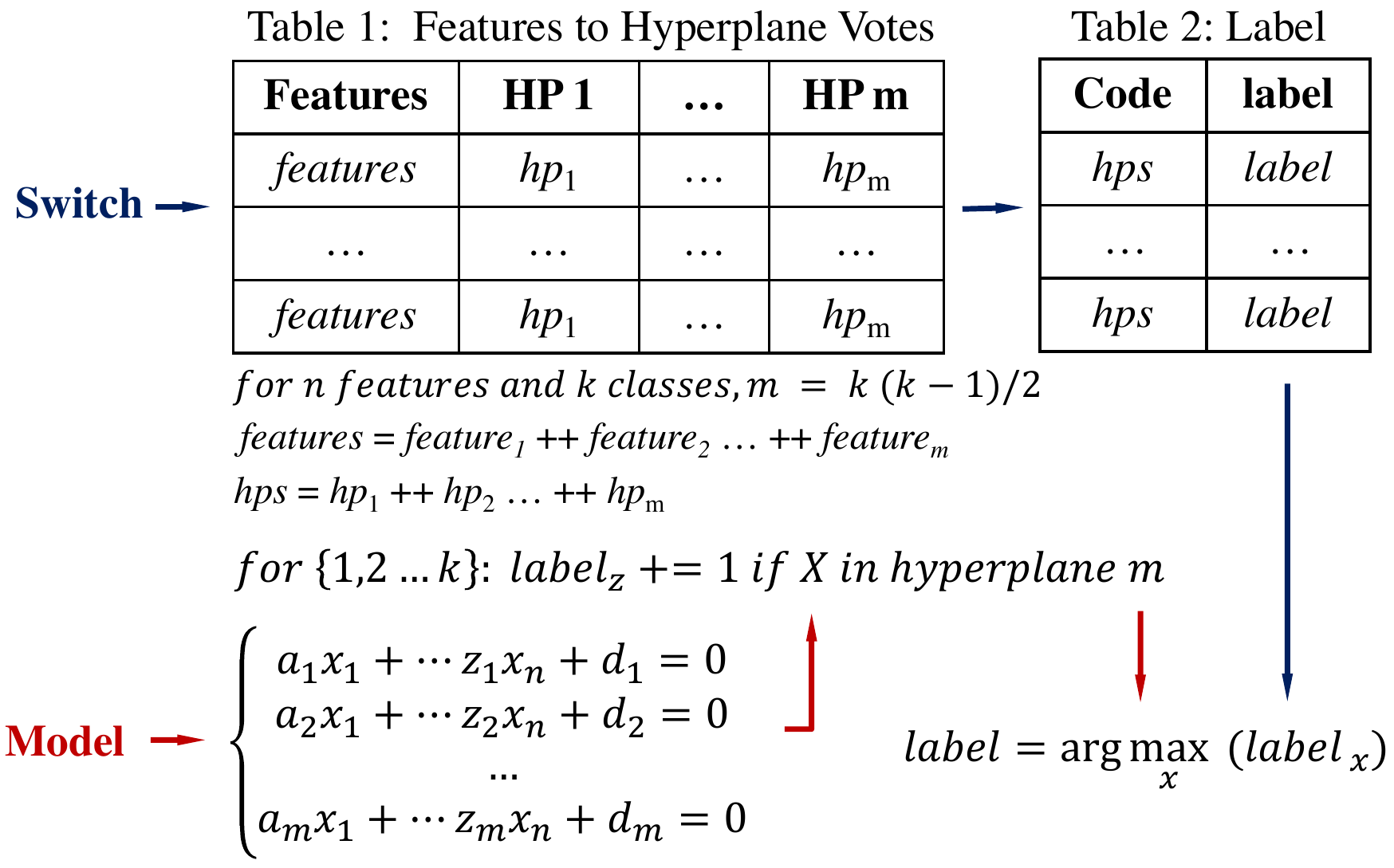}
	\caption{Mapping an SVM Model.}
	\label{fig:model_svm}
	\vspace{-1em}
\end{figure}
A table per hyperplane means that $m$ \lut are used, one per hyperplane and the outcome of the lookup indicates on which side of a hyperplane is a given input. The key to a table is a set of features, and the action is a ``vote''. A ``vote'' is a one-bit value mapped to the metadata bus that indicates if the input belongs within or outside a hyperplane. The ``votes'' from all $m$ tables are counted in the last stage, and the class with the highest count of ``votes'' is the classification result. 

The table per hyperplane approach is feasible only when the concatenation of all features does not lead to a too wide key. 
 If the features used are, for example, source and destination port, protocol, and some IP flags, the key will be relatively small, and the solution will be feasible. Theoretically, the concatenation of all features can yield the classification within a single table.
However, this table is likely to be very large and less resource-efficient than distributing across a few smaller tables.

The main advantage of the table per hyperplane approach is that there is no unintentional loss of accuracy; the output of the table is a vote, not a value.
It is possible to purposely lose some accuracy, e.g., if one wants to reduce the number of table entries by merging multiple ranges of different ``votes'' into a single entry (e.g., if keys 0-1111 and 1113-32767 are mapped to class 1, and key 1112 to class 2). In contrast, a solution using a table per feature may lose some accuracy, as the result of a lookup is a calculated value (and not a code), which has an accuracy limited by its number of bits. 
 The final classification decision may not be affected by the loss of accuracy in calculations along the pipe, but this is not guaranteed. 
 
Using a table per feature will still be favored in some cases, e.g., if working with eight features, each of eight-bit, so each table is only (and at most) 256 entries deep, and features can be looked up in parallel. The table per hyperplane equivalent will be multiple ($m$) tables of a 64-bit key.

\begin{figure}[h]
	\centering
	\includegraphics[width=1\columnwidth]{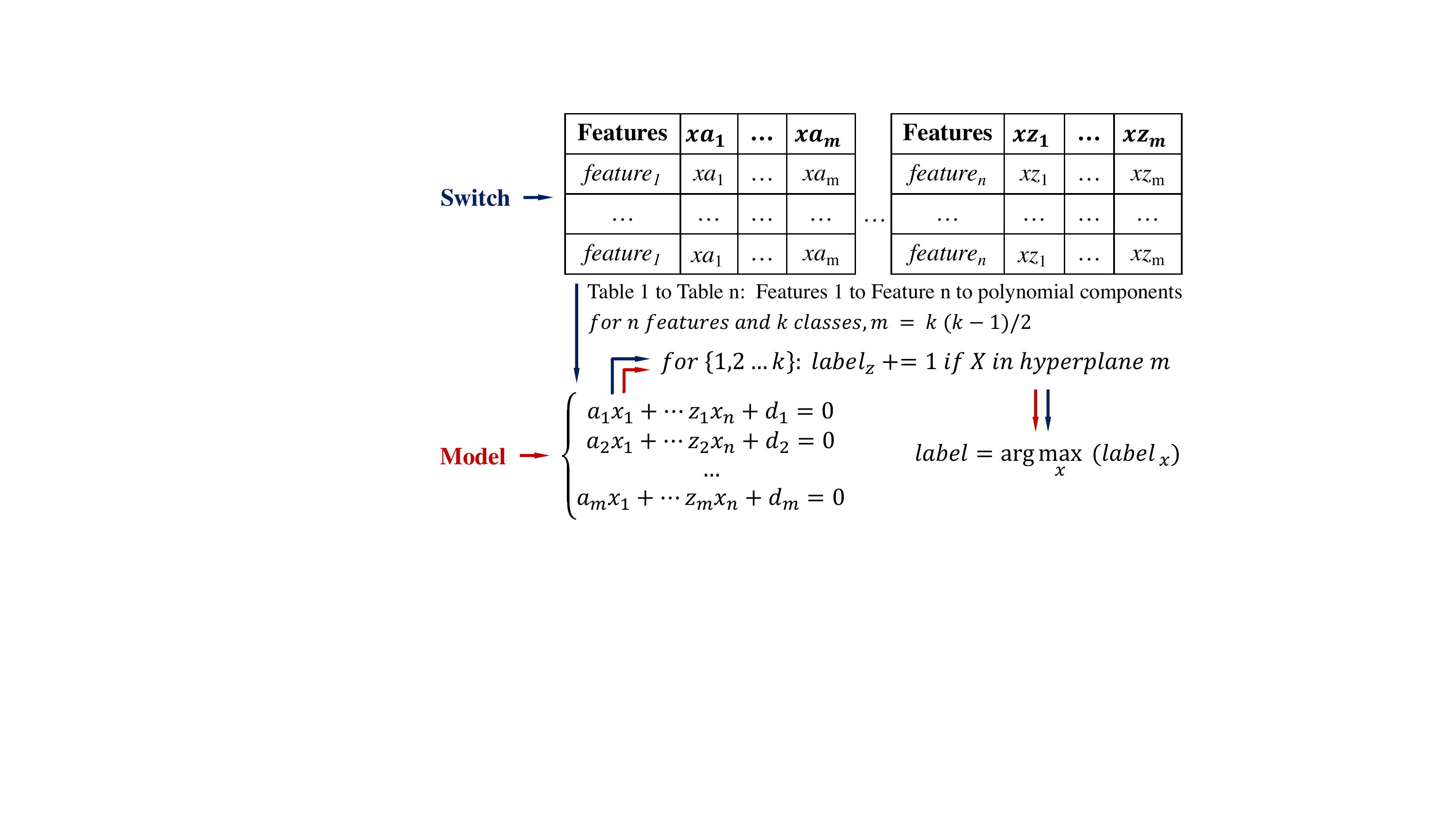}
	\caption{Mapping an SVM Model (Second Option)}
	\label{fig:model_svm2}
\end{figure}

\subsection{Na\"{\i}ve Bayes}\label{sec:bayes}

For a Na\"{\i}ve Bayes classifier~\cite{maron1961automatic}, we assume a Gaussian distribution of independent features~\cite{hand2001idiot}. Similar concepts apply to related methods, such as kernel estimation~\cite{moore2005internet}. Under this assumption, the likelihood of feature $x_i$ is expressed as:
\[P(x_i|y) = \frac{1}{\sqrt{2\pi \sigma^{2}_{y}}}exp\Big(-\frac{(x_{i}-\mu_y)^{2}}{2 \sigma^{2}_{y}}\Big)\]
\noindent And the classification rule is:
\[\hat{y} = arg\:max_y P(y)\prod_{i=1}^nP(x_i|y)\]
If there are $n$ features and $k$ classes, there are $k\times n$ pairs of $(\mu_y,\sigma_{y})$.

 \begin{figure}[h]
 \vspace{2em}
 	\centering
 	\includegraphics[width=1\columnwidth]{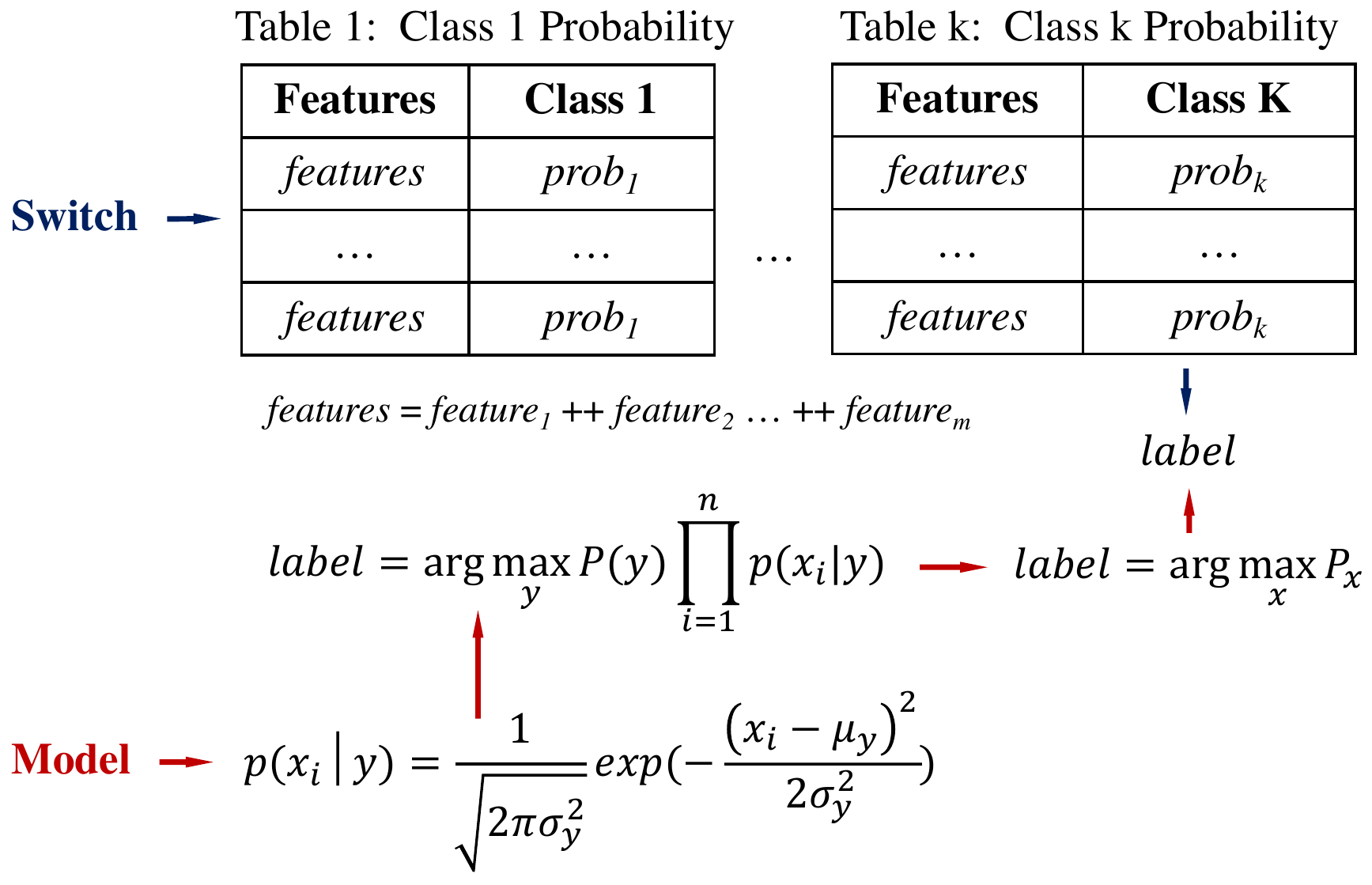}
 	\caption{Mapping a Bayes Model}
 	\label{fig:model_bayes}
 \end{figure}

A mapping based on a table per feature is possible but can be both inefficient and inaccurate. Here, the result of each feature-value lookup will be a vector of probabilities. As the number of bits per vector is limited, there will be some accuracy loss. Even if the target allows for any vector length and a fixed point notation is used, the amount of metadata that needs to be carried between stages will be higher than other solutions, and depending on the number of features and classes, exceeding allowed resources. In this approach, each class will require a table at the end of the pipe to calculate its overall probability, bringing the overall number of tables required to $O(n+k)$. Unless there is a compromise on accuracy, the number of entries in each such table will be large, as the key is the concatenation of all probabilities per class. Finally, at the end of the pipeline, a comparison is required to find  $max_y P(y)$. 

\begin{figure}[h]
	\centering
	\includegraphics[width=1\columnwidth]{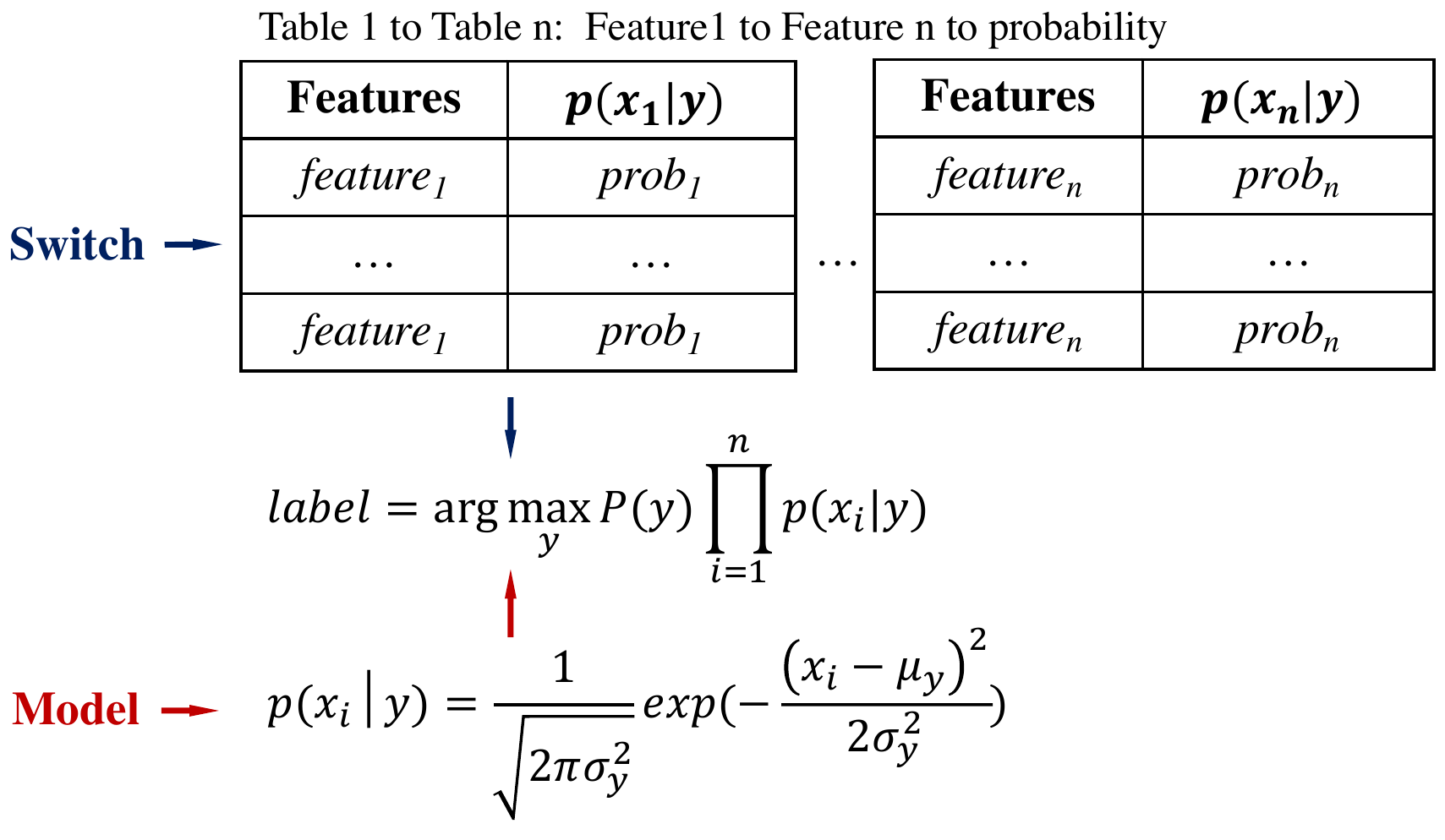}
	\caption{Mapping a Bayes Model (Second Option)}
	\label{fig:model_bayes2}
\end{figure}
\begin{figure*}[htbp]
	\centering
	\includegraphics[width=1\linewidth]{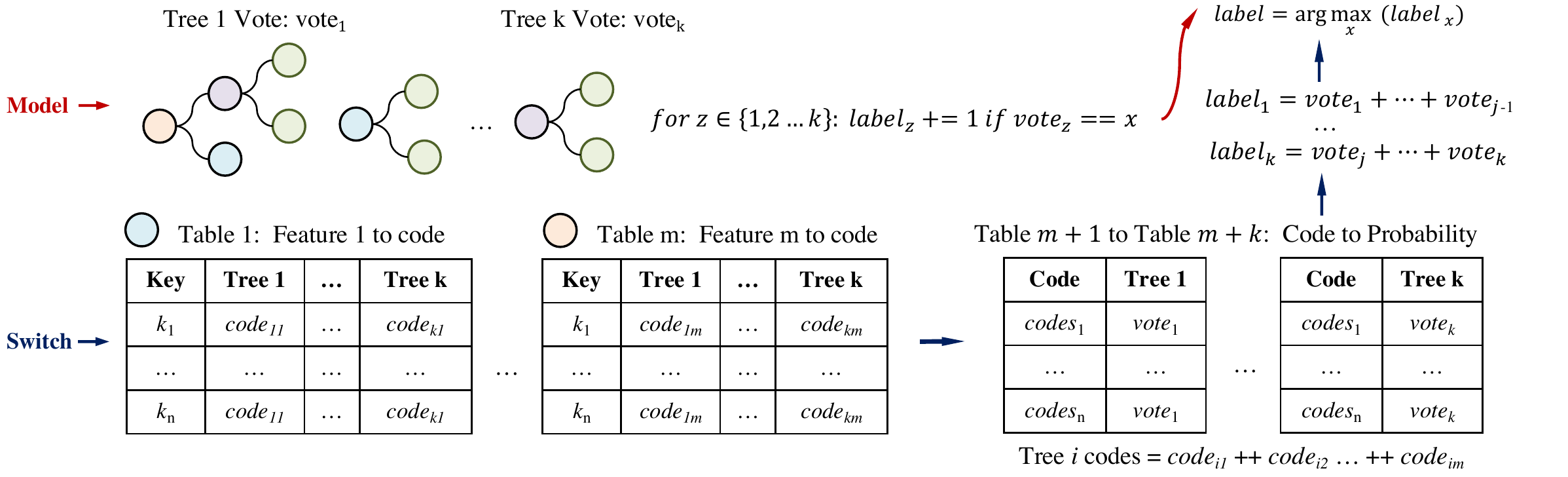}
	\caption{Mapping a Random Forest Model}
	\label{fig:model_rf}
\end{figure*}

\section{Mapping Models - Additional Information}
A better approach is to use one table per class, with all the features as the key, and with the result being the probability of that class. The disadvantage is the size of the required table: it uses a very wide key (a form of a concatenation of all input features values), and its depth is proportional to this width unless a compromise is made for accuracy. The resulting probability does not need to be presented as a fraction, and an integer value can be used that symbolizes the probability. As long as the same notation is used across all tables, the final comparison of  $P(y)$ and the classification result will be correct. Figure~\ref{fig:model_bayes} in Appendix~\ref{app:models} illustrates this implementation.

\subsection{K-Means}\label{sec:kmeans}

An example of unsupervised learning mapped to a network device uses K-means clustering. In K-means, $k$ classes are represented by $k$ centers of clusters, with each center defined by $n$ coordinate values, one per feature. A data point will be mapped to a class based on its nearest center of a cluster. The distance from cluster $i$ is denoted by:
$$D_{i} = \sqrt{(x_{1}-c^{i}_{1})^2+(x_{2}-c^{i}_{2})^2+..(x_{n}-c^{i}_{n})^2}$$
where $x_1$ to $x_n$ are the values of the data point's features. Obviously, to find the nearest cluster, it is sufficient to consider the square distances.

As in previous examples, there are two ways to map the model to a network device. One option is using a table per feature, with the lookup's result of table $i$ being a vector of $\{(x_{i}-c^{1}_{i})^2,(x_{i}-c^{2}_{i})^2,...,(x_{i}-c^{k}_{i})^2\}$. Here, the last stage will need to sum up all $D_{i}$ and find the smallest one\footnote{Values can also be summed up in each stage.}. As before, the challenges here are the accuracy of the calculation and the width of the required metadata bus.

 \begin{figure}[h]
	\centering
	\includegraphics[width=1\columnwidth]{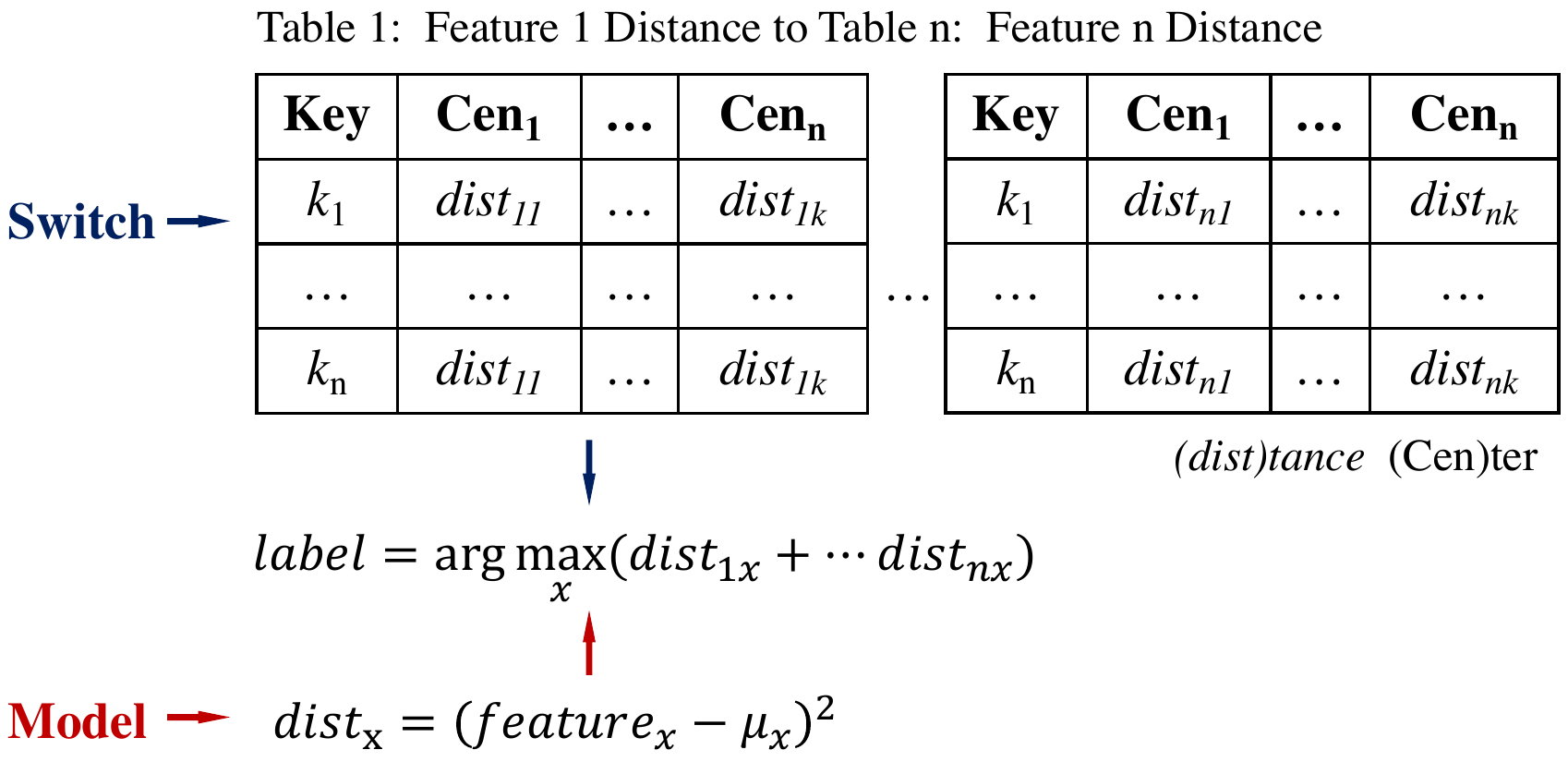}
	\caption{Mapping a K-means Model}
	\label{fig:model_kmeans}
	\vspace{-0.5em}
\end{figure}

The second approach uses a table per class, with the key being the concatenation of all features (presenting a challenge of key width). The result of each such table lookup is the distance of the data point from the center of the cluster. As proposed above, this distance can be consistently represented by an integer value across all tables, allowing for easy comparison and selection at the last stage. The approaches are illustrated in Appendix~\ref{app:models}.

 \begin{figure*}[htbp]
	\centering
	\includegraphics[width=1\linewidth]{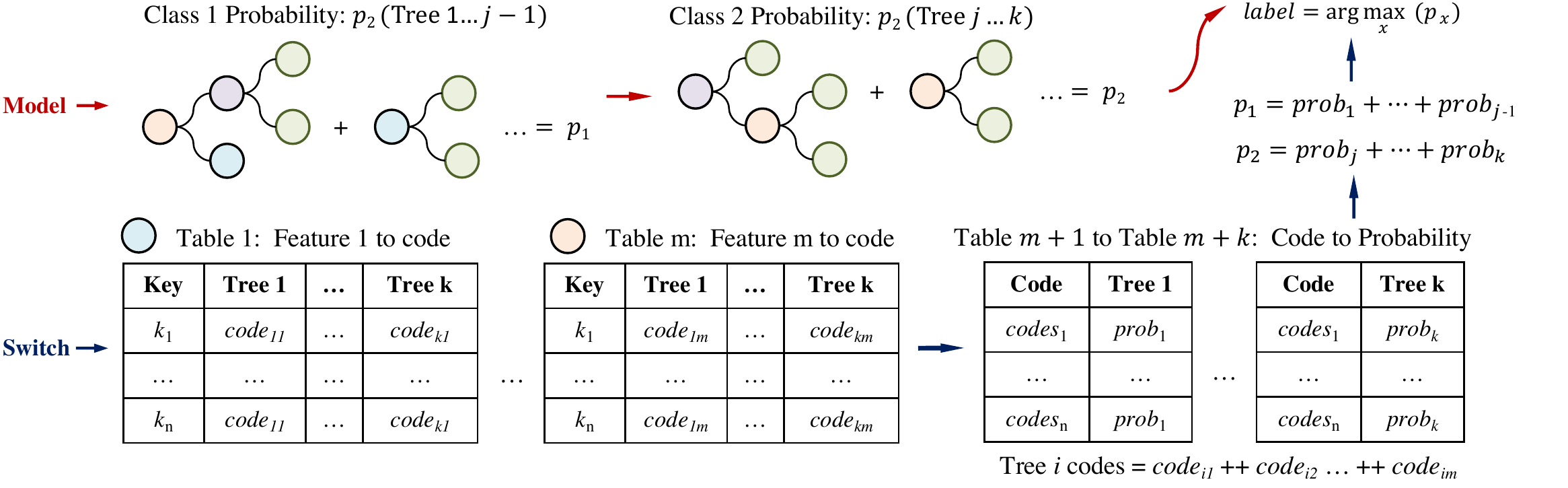}
	\caption{Mapping an XGBoost Model.}
	\label{fig:model_xgboost}
	\vspace{-0.5em}
\end{figure*}

\section{Feature Extraction on NetFPGA}\label{sec:app_netfpga}

Using our implementation on NetFPGA, we explore the feasibility of implementing the full range of features included in the UNSW dataset~\cite{unsw-dataset}, regardless of their contribution to the Random Forest model's performance.

The dataset includes 47 features, which we categorize into three groups according to their implementation complexity, with a fourth group including with unclear description. We classify 11 features as easy, 15 features as medium complexity and 18 features as hard or impossible to implement on a switch. Three more features are unclear. Out of these, we implement 15 features on NetFPGA.

In general, easy features are stateless, though some require operating on the data, e.g., comparison. 
All the features implemented under this class are packet-level features.

The medium level features are stateful, and include both counted and time related flow level features. Among the features that we implement under this category we include flow duration, flow record start time and last time, packet inter-arrival time, packets loss per flow, and data rate (bits per second).  For the data rate, we use both flow duration and a counter of bytes per flow, and estimate data rate by the ratio of byte per flow to duration.  
The most complex feature implemented is jitter, where we use an multiple bins to store packets inter arrival time per flow, and consider the number of packet in each bin according to the classifier.

The hard to impossible features require tracking state machines, and information that is not available on the switch. 

The implementation that supports all features uses 13 externs in total, where some of the extrens are used by more than one feature (with a single access in the pipeline).
The types of externs use are hash, read/write memory (equivalent to registers) and read-modify-write memory (atomic operation). The types of externs available on NetFPGA are described in ~\cite{ibanez2019p4netfpga}.

Extracting a jitter feature illustrates the complexity of  some flow-level features. The jitter feature, in the context of this example , considers the difference in inter-arrival time between packets of the same flow. As a new packet arrives, its flow-identifier fields are extracted in the Parser, and an extern (e.g., timestamp mechanism) is used to get the arrival time. A hashing function turns the flow identifier fields into a flow-id. These steps are common in non-\ml use cases. Next, the previous arrival time in the flow is read from a memory, and replaced with the new timestamp (e.g., using read-modify-write). 
The difference between the current and the previous arrival time is the inter-arrival gap. To account for the jitter, bins can be used, e.g., number of packets with less than 1ms, 1ms to 10ms, or more than 10ms inter-arrival time. Each bin requires a per-flow entry in the memory. 
Given $K$ flows and $N$ bins means that the jitter feature will required $K\times(N+1)$ memory entries.

\section{Related Work - Extension}\label{sec:app_related}

\begin{figure}[htb]
	\centering
	\includegraphics[width=1\columnwidth]{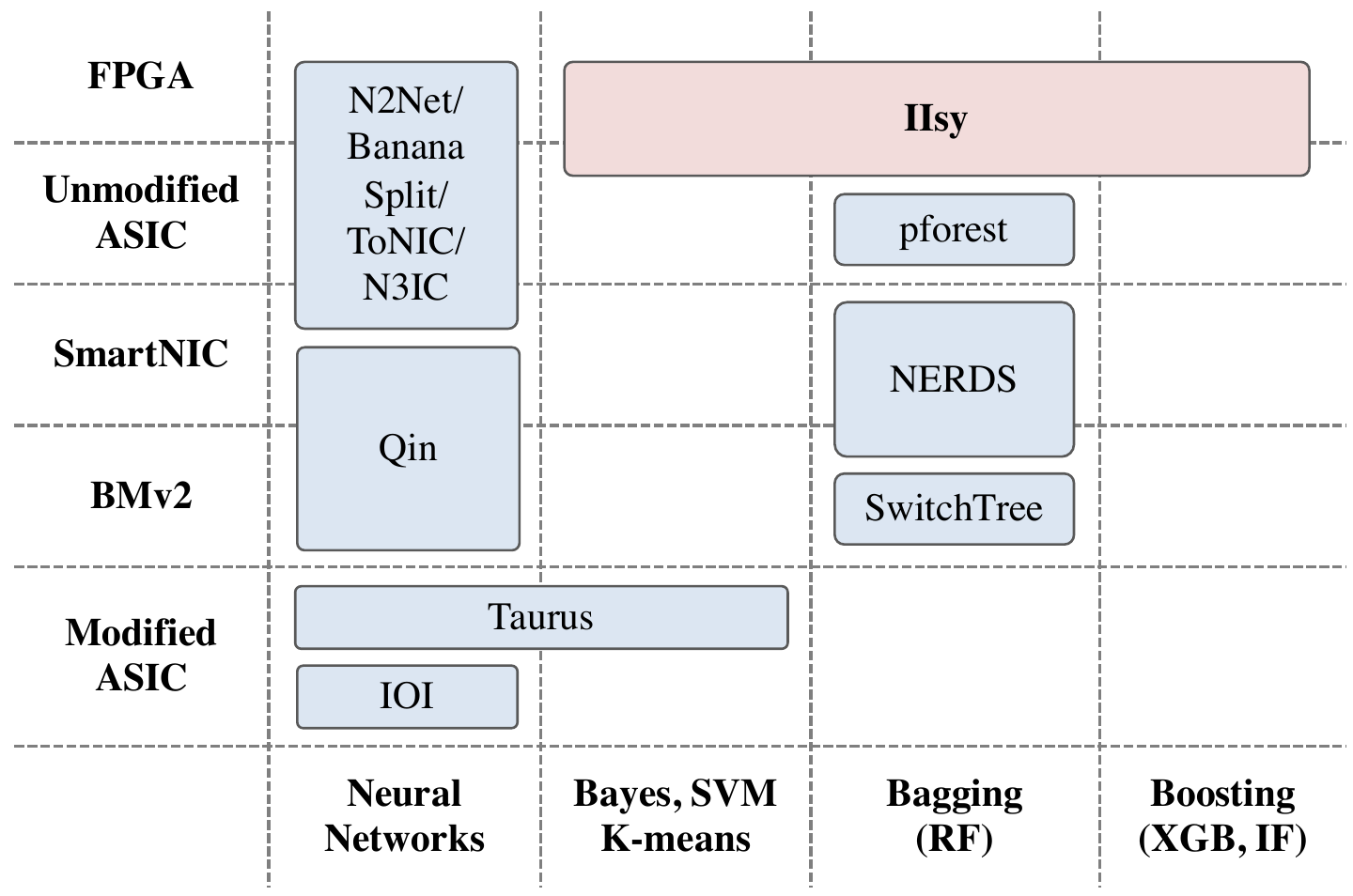}
	\vspace{-1em}
	\caption{The positioning of IIsy}
	\label{fig:positioning2}
	\vspace{-1em}
\end{figure}

Figure~\ref{fig:positioning2} illustrates the positioning of IIsy's contribution relative to the works listed in Table~\ref{tab:related_comparison}. As the figure shows, IIsy is the only work to present a generic solution for a range of \ml methods. 

While multiple works have considered Random Forest, they focused on bmv2 and smart NICs. pForest has also focused on bmv2, and presented a non-optimized implementation on Tofino with a depth of 4, significantly less than IIsy. Solutions such as~\cite{xavier2021programmable,lee2020switchtree} did not attend to resource constraints, and don't scale as well as IIsy. Even in the bmv2 implementation, SwitchTree~\cite{lee2020switchtree} was studied with only 5 trees and depth of 10, where each tree is coded independently.  

While Taurus supports SVM and KMeans, it is not possible to compare to it, as it did not report \ml performance results for these models. Moreover, it relies on a modification to the silicon design.
 
\section{Test Setup}\label{sec:app_setup}
Our system test environment uses APS-Networks BF6064X, an Intel Barefoot Tofino platform with $64\times100G$ ports. Barefoot's SDE 9.2.0 is used on the switch, and we further experiment with SDE 9.6.0 in the software development environment.
P4-NetFPGA~\cite{ibanez2019p4netfpga} with SDNet 2018.2 compiler is used for the FPGA development. 

ESC4000A-E10 servers using AMD EPYC 7302P  CPUs with 256GB RAM, Ubuntu 20.04LTS, and equipped with Mellanox ConnectX-5 100G NICs are used to send traffic to the switch using DPDK 20.11.1 and PktGen 21.03.0. Four CPU cores are dedicated per port. 

To test full throughput, we use a snake configuration, where traffic is looped from each port to the following one, enabling traffic across all 64 ports, which is a common practice~\cite{dang2020p4xos}. A set of python scripts is used to generate, capture and check traffic. As a baseline, we measure 6.2Tbps on the switch when running simple forwarding.


In this section we include additional illustrations of mapping different machine learning models to network devices. The methodology of mapping these models is described in~\ref{sec:mapping}.


\flushend

\end{document}